\documentclass[%
 twocolumn,
 superscriptaddress,
 aps,
 prb,
 floatfix,
]{revtex4-2}

\usepackage{import}
\usepackage{graphicx}	
\usepackage{dcolumn}	
\usepackage{bm}		
\usepackage{hyperref}	

\usepackage{graphicx}
\usepackage{dcolumn}
\usepackage{amsmath}
\usepackage{color}
\usepackage{bbold}
\usepackage{color}
\usepackage{natbib}

\newcommand{\RN}[1]{\textup{\uppercase\expandafter{\romannumeral#1}}}%

\begin{document}

\title{Influence of pressure on properties of multi-gap type-I superconductor BeAu}

\author{Rustem Khasanov}
\email[]{rustem.khasanov@psi.ch}
\affiliation{Laboratory for Muon Spin Spectroscopy, Paul Scherrer Institute, 5232 Villigen, Switzerland}

\author{Riccardo Vocaturo}
\email[]{r.vocaturo@ifw-dresden.de}
\affiliation{Leibniz Institute for Solid State and Materials Research (IFW) Dresden, Helmholtzstrasse 20, 01069 Dresden, Germany}

\author{Oleg Janson}
\affiliation{Leibniz Institute for Solid State and Materials Research (IFW) Dresden, Helmholtzstrasse 20, 01069 Dresden, Germany}

\author{Andreas Koitzsch}
\affiliation{Leibniz Institute for Solid State and Materials Research (IFW) Dresden, Helmholtzstrasse 20, 01069 Dresden, Germany}

\author{Ritu Gupta}
\affiliation{Laboratory for Muon Spin Spectroscopy, Paul Scherrer Institute, 5232 Villigen, Switzerland}
\affiliation{Department of Physics, Indian Institute of Technology Ropar, Rupnagar, Punjab 140001, India}

\author{Debarchan Das}
\affiliation{Laboratory for Muon Spin Spectroscopy, Paul Scherrer Institute, 5232 Villigen, Switzerland}

\author{Nicola P.M. Casati}
\affiliation{Laboratory for Synchrotron Radiation – Condensed Matter, Paul Scherrer Institut, Villigen PSI CH-5232, Switzerland}

\author{Maia G. Vergniory}
\affiliation{Max-Planck-Institut für Chemische Physik fester Stoffe, Nöthnitzer Straße 40, 01187 Dresden, Germany}

\author{Jeroen van den Brink}
\email[]{j.van.den.brink@ifw-dresden.de}
\affiliation{Leibniz Institute for Solid State and Materials Research (IFW) Dresden, Helmholtzstrasse 20, 01069 Dresden, Germany}

\author{Eteri Svanidze}
\email[]{svanidze@cpfs.mpg.de}
\affiliation{Max-Planck-Institut für Chemische Physik fester Stoffe, Nöthnitzer Straße 40, 01187 Dresden, Germany}

\begin{abstract}
We report on studies of the superconducting and normal state properties of the noncentrosymmetric superconductor BeAu under hydrostatic pressure conditions. The room-temperature equation of state (EOS) reveals the values of the bulk modulus ($B_0$) and its first derivative ($B^\prime_0$) at ambient pressure to be $B_0 \simeq 132$~GPa and $B^\prime_0 \simeq 30$, respectively. Up to the highest pressures studied ($p \simeq 2.2$~GPa), BeAu remains a multi-gap type-I superconductor. The analysis of $B_{\rm c}(T, p)$ data within the self-consistent two-gap approach suggests the presence of two superconducting energy gaps, with the gap-to-$T_{\rm c}$ ratios $\Delta_1/k_{\rm B}T_{\rm c} \sim 2.3$ and $\Delta_2/k_{\rm B}T_{\rm c} \sim 1.1$ for the larger and smaller gaps, respectively [$\Delta = \Delta(0)$ is the zero-temperature value of the gap and $k_{\rm B}$ is the Boltzmann constant]. With increasing pressure, $\Delta_1/k_{\rm B}T_{\rm c}$ increases while $\Delta_2/k_{\rm B}T_{\rm c}$ decreases, suggesting that pressure enhances (weakens) the coupling strength between the superconducting carriers within the bands where the larger (smaller) superconducting energy gap has opened. The superconducting transition temperature $T_{\rm c}$, \textcolor{black}{the zero-temperature values of the superconducting gaps $\Delta_1$ and $\Delta_2$} and the zero-temperature value of the thermodynamic critical field $B_{\rm c}(0)$ decrease with increasing pressure, with the rates of ${\rm d}T_{\rm c}/{\rm d}p \simeq -0.195$~K/GPa, \textcolor{black}{${\rm d}\Delta_1/{\rm d}p \simeq -0.034$~meV/GPa, ${\rm d}\Delta_2/{\rm d}p \simeq -0.029$~meV/GPa,} and ${\rm d}B_{\rm c}(0)/{\rm d}p = -2.65(1)$~mT/GPa, respectively. The measured $B_{\rm c}(0)$ values plotted as a function of $T_{\rm c}$ follow an empirical scaling relation established for conventional type-I superconductors.
\end{abstract}

\maketitle

\section{Introduction}

Since the discovery of superconductivity over a century ago, significant progress was made in both utilizing and understanding this fascinating phenomenon. Numerous groups of materials exhibiting superconductivity were discovered and studied thoroughly, with the aim of uncovering mechanisms for electron pairing that could lead to higher critical temperatures and fields.

Noncentrosymmetric superconductors, which lack inversion symmetry in their crystal structure, exhibit a variety of unusual properties \cite{Sigrist_AIP-Cof-Proc_2009, Yip_AnRevCondMat_2014, Smidman_RPP_2017}. The absence of inversion symmetry results in Fermi surface splitting, which promotes both intraband and interband pairing \cite{Takimoto_JPSJ_2009}. In some cases, in addition to inversion, also mirror symmetries are absent from the crystal structure. In this case, superconductivity of such (chiral) systems may exhibit an axion electromagnetic response that, for example, gives rise to a chiral Meissner state, where a magnetic field penetrating into the bulk starts to steadily rotate away from the applied field \cite{Shyta}.

\begin{figure}[b]
\includegraphics[width=\linewidth]{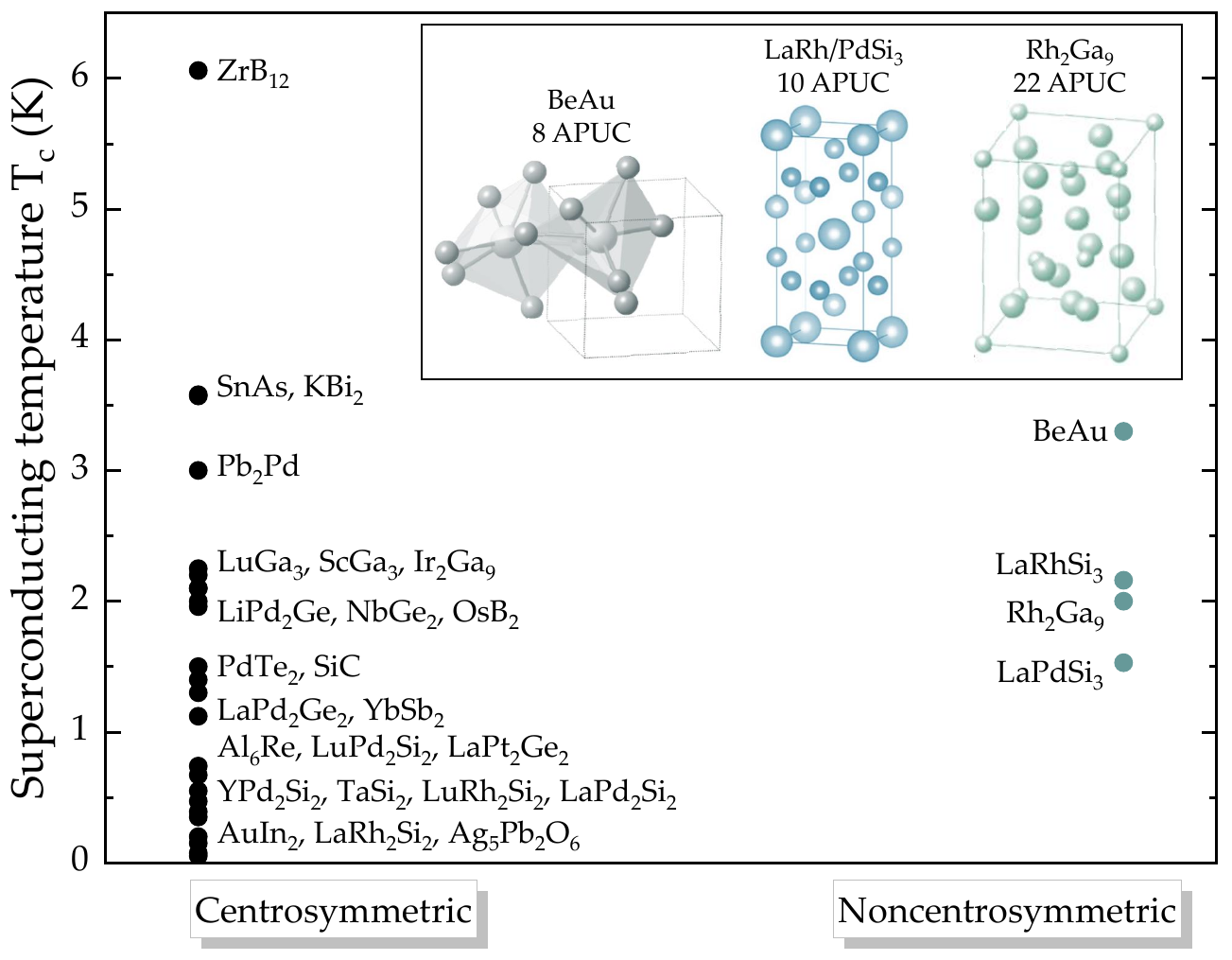}
\caption{Centrosymmetric (left) and noncentrosymmetric (right) type-I superconductors discovered to date. Among noncentrosymmetric superconductors, BeAu displays the highest values of the critical temperature $T_{\rm c}$ and the simplest structures with only 8 atoms per unit cell (APUC).}
 \label{Type_I_SCs}
\end{figure}

BeAu has emerged as a notable noncentrosymmetric superconductor with a chiral structure, exhibiting a transition temperature $T_{\rm c} \simeq 3.2$~K \cite{Matthis_JPCS_1959, Rebar_thesis_2015, Amon_PRB_2018, Singh_PRB_2019, Beare_PRB_2019, Rebar_PRB_2019}. It is characterized by a multiband electronic structure, with several energy bands crossing the Fermi level \cite{Rebar_thesis_2015, Rebar_PRB_2019}. This multiband nature suggests the possibility of complex superconducting behaviors, as multiple bands contribute differently to the superconducting condensate.

Experimental investigations and band-structure calculations have provided compelling evidence for multiband superconductivity in BeAu \cite{Rebar_thesis_2015, Rebar_PRB_2019, Khasanov_AuBe_PRR_2020, Khasanov_AuBe_PRB_2020, Datta_PRB_2022}. In particular, the specific heat measurements and thermodynamic critical field data analysis strongly support the presence of two superconducting contributions \cite{Khasanov_AuBe_PRR_2020, Khasanov_AuBe_PRB_2020}. Further confirmation came from tunneling experiments revealing the presence of two superconducting energy gaps with $\Delta_1/k_{\rm B}T_{\rm c} \sim 2.2$ and $\Delta_2/k_{\rm B}T_{\rm c} \sim 1.2$ for the larger and smaller gaps, respectively [$\Delta = \Delta(0)$ is the zero-temperature value of the gap and $k_{\rm B}$ is the Boltzmann constant] \cite{Datta_PRB_2022}. However, two subsequent muon-spin rotation/relaxation ($\mu$SR) experiments have not confirmed the presence of TRSB in BeAu \cite{Singh_PRB_2019, Beare_PRB_2019}. The absence of TRSB and the presence of multiple contributions to the superconducting energy gap indicate that while BeAu's superconductivity is likely phonon-mediated, it may exhibit some of the more exotic properties associated with unconventional noncentrosymmetric superconductors. Another interesting feature of BeAu is that it is a type-I superconductor, which is fairly rare among compounds \cite{Peets, lv2020a, svanidze2012a, zhao2012a, takeda2015a, biswas2020a, anand2011a, smidman2014a, yonezawa2005a, hull1981a, palstra1986a, Leng_PRB_2017, g2020a, nakamura2013a, herrmannsdoerfer1996a, kobayashi1981a, sun2016a, bekaert2016a, Arushi_PRB_2021, ren2007a, wang2014a}. Furthermore, as summarized in Figure~\ref{Type_I_SCs}, only a handful of type-I superconductors have noncentrosymmetric crystal structures. So far, these materials appear to form with fairly simple structures, as evidenced by a low number of atoms per unit cell (APUC).

Tuning of superconductivity in BeAu remains an area of significant interest. Potential methods include doping, introducing various magnetic or non-magnetic impurities, or applying uniaxial or hydrostatic pressures. In our study, we use hydrostatic pressure since it is considered as a cleaner method compared to impurity or doping techniques as it maintains a nearly similar crystal structure (without pressure-induced structural transitions) and does not introduce additional scattering centers. In superconducting materials, moderate pressures are typically enough to affect the phonon spectra as well as electron-phonon coupling constants \cite{Lorenz_2005, Schilling_book_2007, Schilling_JPCS_2008}. Our findings provide new insights into the superconducting properties of BeAu and indicate the robustness of its type-I superconductivity under varying pressure conditions.

\textcolor{black}{The paper is organized as follows. Section~\ref{sec:ExpDetails} describes the sample preparation procedure and details of the x-ray, AC susceptibility, and $\mu$SR under pressure experiments. The determination of the equation of state of BeAu at $T \simeq 300$~K is reported in Sec.~\ref{seq:EOS}. The pressure dependence of the superconducting transition temperature $T_{\rm c}$ is described in Sec.~\ref{sec:ACS}. Calculations of the effect of pressure on the electron and phonon density of states are given in Sec.~\ref{sec:DFT-DOS}.
The results of $\mu$SR measurements and the determination of the pressure-temperature evolution of the thermodynamic critical field $B_{\rm c}(T,p)$ are summarized in Sec.~\ref{sec:Bc}. The linear scaling between $T_{\rm c}$ and the zero-temperature values of the thermodynamic critical field $B_{\rm c}(0)$ is discussed in Sec.~\ref{sec:Bc_vs_Tc}. Conclusions follow in Sec.~\ref{sec:conclusions}.}

\section{Experimental and theoretical details}\label{sec:ExpDetails}

The polycrystalline BeAu samples were synthesized by arc melting from elements Be (Heraeus, $>99.9$ wt.\%) and Au (Alfa Aesar, $>99.95$ wt.\%) in a 51:49 ratio. A small excess of beryllium was added to compensate for the Be loss due to evaporation. Complete sample preparation was performed in argon-filled glove boxes [MBraun, $p({\rm H}_2{\rm O/O}_2) < 0.1$~ppm], dedicated to the handling of Be-containing samples \cite{Leithe-Jasper_MPI-report_2003}. The as-cast samples were placed in an alumina crucible, sealed in a tantalum tube, and annealed in a tube furnace (HTM Reetz, Berlin, Germany) at 400$^{\circ}$C for 48~h in an inert Ar atmosphere.

Pressure-dependent lattice parameters of BeAu at room temperature were obtained from Pawley fits of synchrotron powder X-ray diffraction data using the software Topas 7 \cite{Topas}. The respective diffraction experiments were carried out at the Materials Science (MS) beamline at the Swiss Light Source (SLS) \cite{Willmott_JSyncRad_2013} using a wavelength of 0.56389 Å, as calibrated from a LaB$_6$ NIST standard using the high resolution diffractometer. For the experiment, a membrane-driven diamond anvil cell (DAC) was used, having diamonds with a 0.5 mm culet, and data collected on a Pilatus 6M detector. The pressure was calibrated by diffraction, using an internal standard of CaF$_2$ mixed with the sample and its equation of state \cite{Angel_1993}. The pressure transmitting medium was Daphne 7575 oil \cite{Daphne}.

\begin{figure}[b!]
\includegraphics[width=1.0\linewidth]{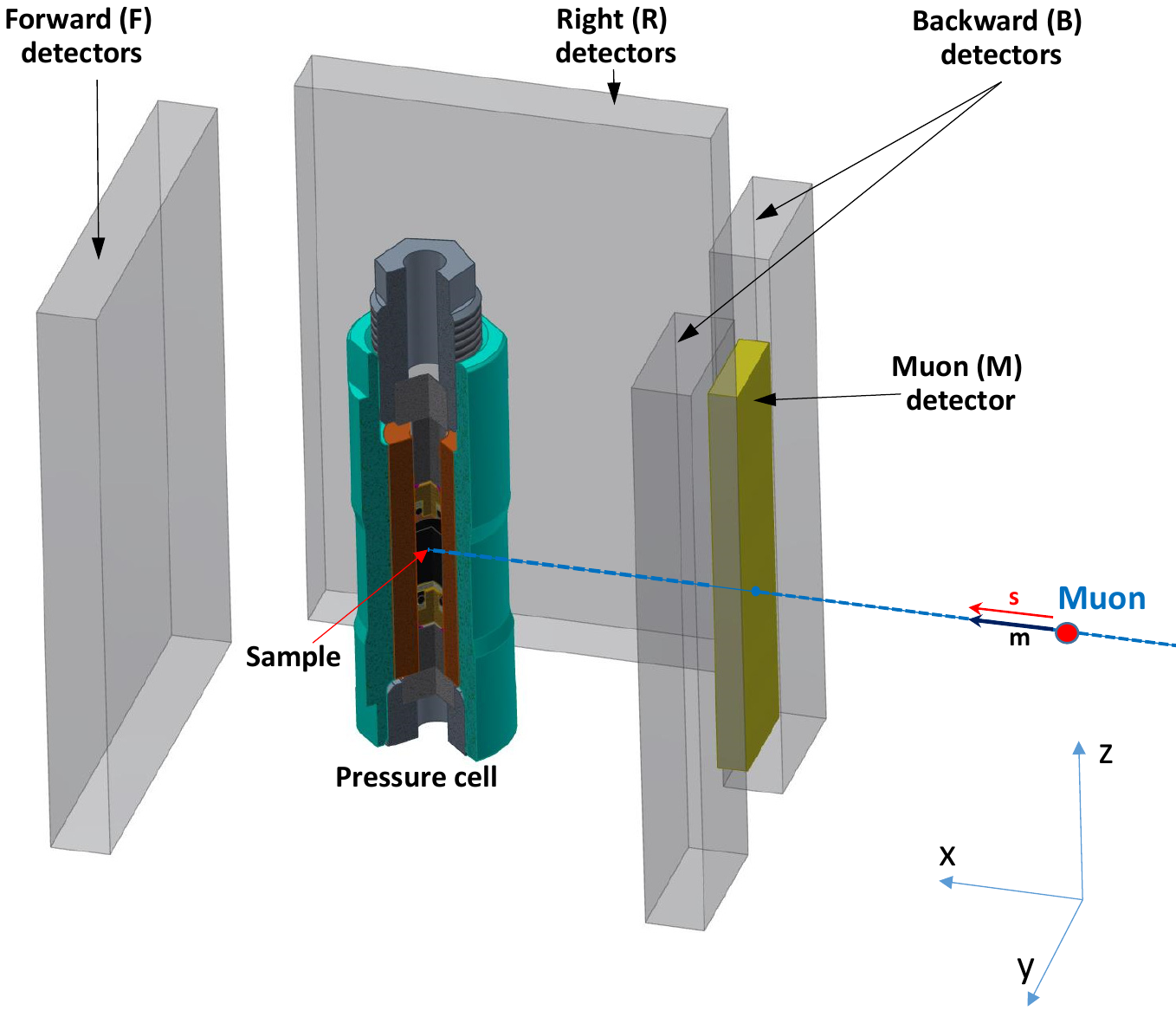}
\caption{\textcolor{black}{The experimental $\mu$SR setup: The sample inside the pressure cell is placed within the detector block, which consists of 11 positron and 1 muon detector (see Ref.~\onlinecite{Khasanov_JAP_2022} for further details). The detectors are coupled to form Backward (B), Forward (F), Left (L), and Right (R) positron counters (the left counter is not shown in the drawing). The external magnetic field ($B_{\rm ap}$) is applied along the y-axis, {\it i.e.}, perpendicular to the initial muon-spin direction ($\bf{s}$).}}
 \label{fig:GPD-setup}
\end{figure}

The $\mu$SR under pressure experiments were conducted at the $\mu$E1 beamline using the General Purpose Decay (GPD) spectrometer \cite{Khasanov_HPR_2016, Khasanov_JAP_2022}. Pressures up to $p \simeq 2.3$~GPa were generated in a double-walled clamp-type cell made of nonmagnetic MP35N alloy \cite{Khasanov_HPR_2016}. Daphne 7373 oil was used as the pressure transmitting medium. Around 20 elliptically shaped disks of BeAu (0.5 mm thick and roughly $3.0 \times 4.0$~mm$^2$ in size) were placed inside the pressure cell channel. The experiments were performed using a $^4$He cryostat equipped with a $^3$He inset in the temperature range 0.25--5~K.

\textcolor{black}{
The experimental $\mu$SR setup is shown schematically in Fig.~\ref{fig:GPD-setup}. The sample, placed inside the double-wall pressure cell, is positioned within the detector block. The detector block consists of 11 positron and 1 muon detectors. Detectors can be coupled together, allowing for the zero-field, transverse-field (TF), and longitudinal field $\mu$SR experiments (see the description of the GPD instrument and various detector setups in Ref.~\onlinecite{Khasanov_JAP_2022}). In the experiments presented here, detectors were combined to form Backward (B), Forward (F), Left (L), and Right (R) positron counters (the left counter is not shown in Fig.~\ref{fig:GPD-setup}). The external magnetic field ($B_{\rm ap}$) was applied parallel to the y-axis, {\it i.e.}, perpendicular to the initial muon-spin ($ \bf{s}$) and the muon momentum ($\bf{m}$) (both are parallel to the x-axis, see Fig.~\ref{fig:GPD-setup}), which corresponds to the transverse field $\mu$SR geometry. The positron-decay spectra were accumulated at Backward (B) and Forward (F) positron detectors. The time evolution of the muon-spin polarization, $P(t)$, was further obtained by using the asymmetry function:
\begin{equation}
A(t) = A_0 P(t) = \frac{N_{\rm F}(t) - \alpha N_{\rm B}(t)}{N_{\rm F}(t) + \alpha N_{\rm B}(t)}.
 \label{eq:asymmetry}
\end{equation}
Here, $A_0$ is the initial asymmetry of the muon-spin ensemble, $N_{\rm F}(t)$ [$ N_{\rm B}(t)$] is the time evolution of the positron counts at the Forward (Backward) detector, and $\alpha$ is a parameter accounting for the different solid angles and efficiencies of the positron detectors. The experimental data were analyzed using the MUSRFIT package \cite{MUSRFIT}.
}

The AC-susceptibility (ACS) experiments were performed using a homemade setup allowing for the simultaneous detection of the superconducting response of the BeAu sample and a small piece of Sn (pressure indicator) placed inside the pressure cell channel along with the main sample. The coil assembly is described in Refs.~\onlinecite{Khasanov_HPR_2016, Naumov_PRA_2022}. A coil surrounding the pressure cell (input coil) is fed with an AC current. Two pickup coils (one centered at the position of the sample and pressure probe) are wound in opposite directions. The signal of the pickup coil is compensated above the superconducting transition temperature $T_{\rm c}$, but it becomes uncompensated when the superconducting transition is crossed upon lowering the temperature. The lack of compensation is detected using a lock-in amplifier (AMETEK 7241) which compares the phase and amplitudes of the input signal with those of the pickup signal.

The photoemission data were measured using a NanoESCA system (Scienta Omicron) equipped with a monochromatized Al K$\alpha$ excitation source (h$\nu$=1486.6 eV) at $T = 300$ K.

The electronic structure of BeAu, was investigated by performing full-relativistic density functional theory (DFT) calculations using the full-potential local-orbital code FPLO \cite{FPLO} (version 22.01-63). For the exchange-correlation term, Perdew-Burke-Ernzerhof \cite{PBE} (PBE) parametrization of the generalized gradient approximation (GGA) was used. For the structural input, the experimental lattice parameters ($P2_13$ space group, $a = 4.6692(2)$ \AA \ \cite{Amon_PRB_2018}) and optimized the internal atomic coordinates with respect to the GGA total energy were considered. The electronic density of states (DOS) were calculated on a 50$\times$50$\times$50 $k$-mesh.

Additionally, the density functional perturbation theory (DFPT) calculations were performed to obtain phonon dispersions and the corresponding phonon DOS. For these calculations,  Quantum Espresso version 6.7 \cite{QE,QE2} with ultrasoft scalar-relativistic pseudopotentials from \url{http://www.quantum-espresso.org} was used. Self-consistent calculations were done on a mesh of 18$\times$18$\times$18 $k$-points, with a smearing of 0.005 Ry, and a cut-off of 120 Ry. Phonon calculations were conducted on a 2$\times$2$\times$2 $q$-grid.

\begin{figure}[htb]
\includegraphics[width=0.9\linewidth]{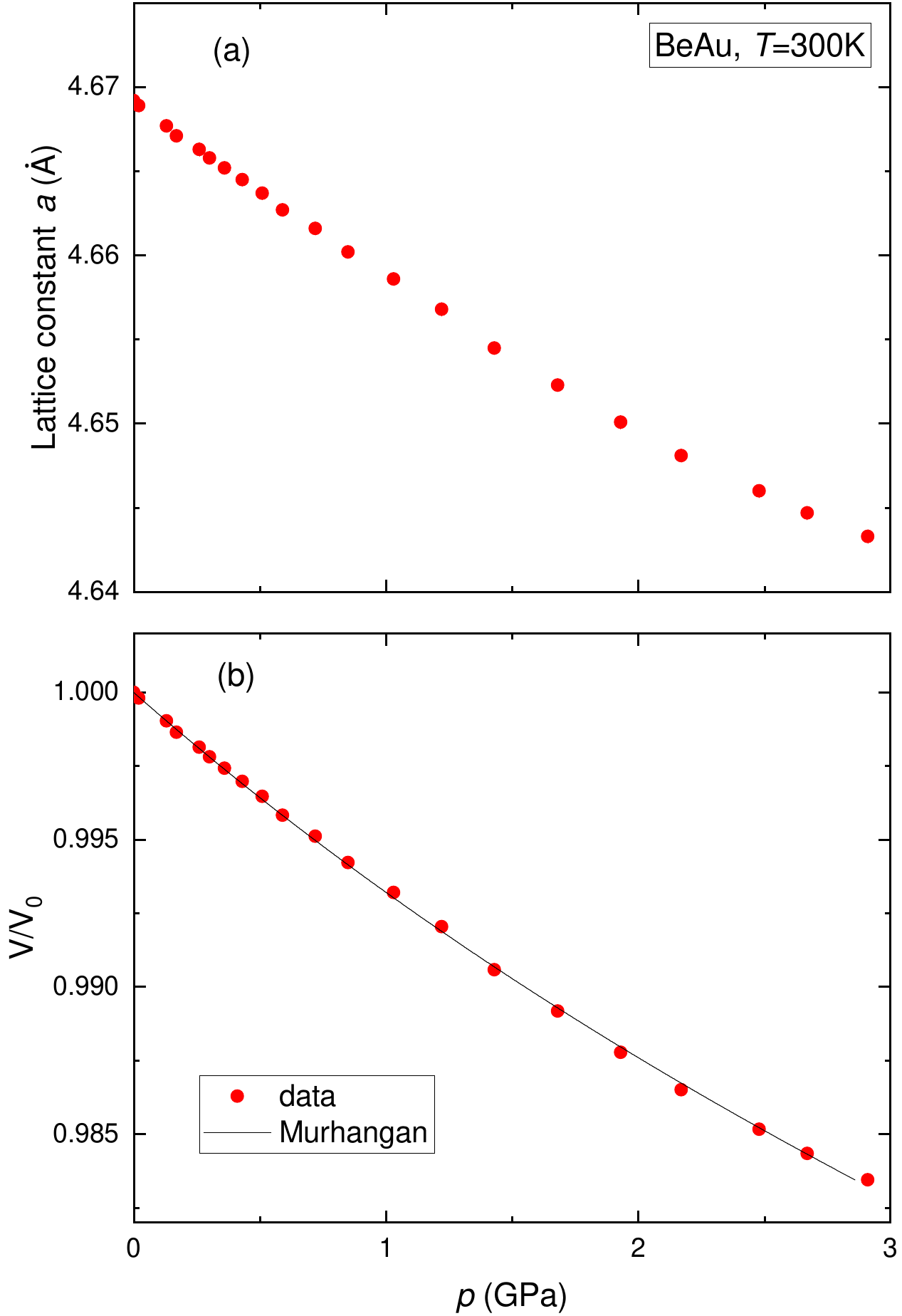}
\caption{(a) Dependence of the lattice constant $a$ on pressure at $T\simeq300$~K. (b) Dependence of $V/V_0$ on pressure. The solid line is the fit \textcolor{black}{with Murnaghan EOS model, Eq.~\ref{eq:Murnaghan}}.}
 \label{fig:Structure}
\end{figure}

\begin{figure*}
\includegraphics[width=0.7\linewidth]{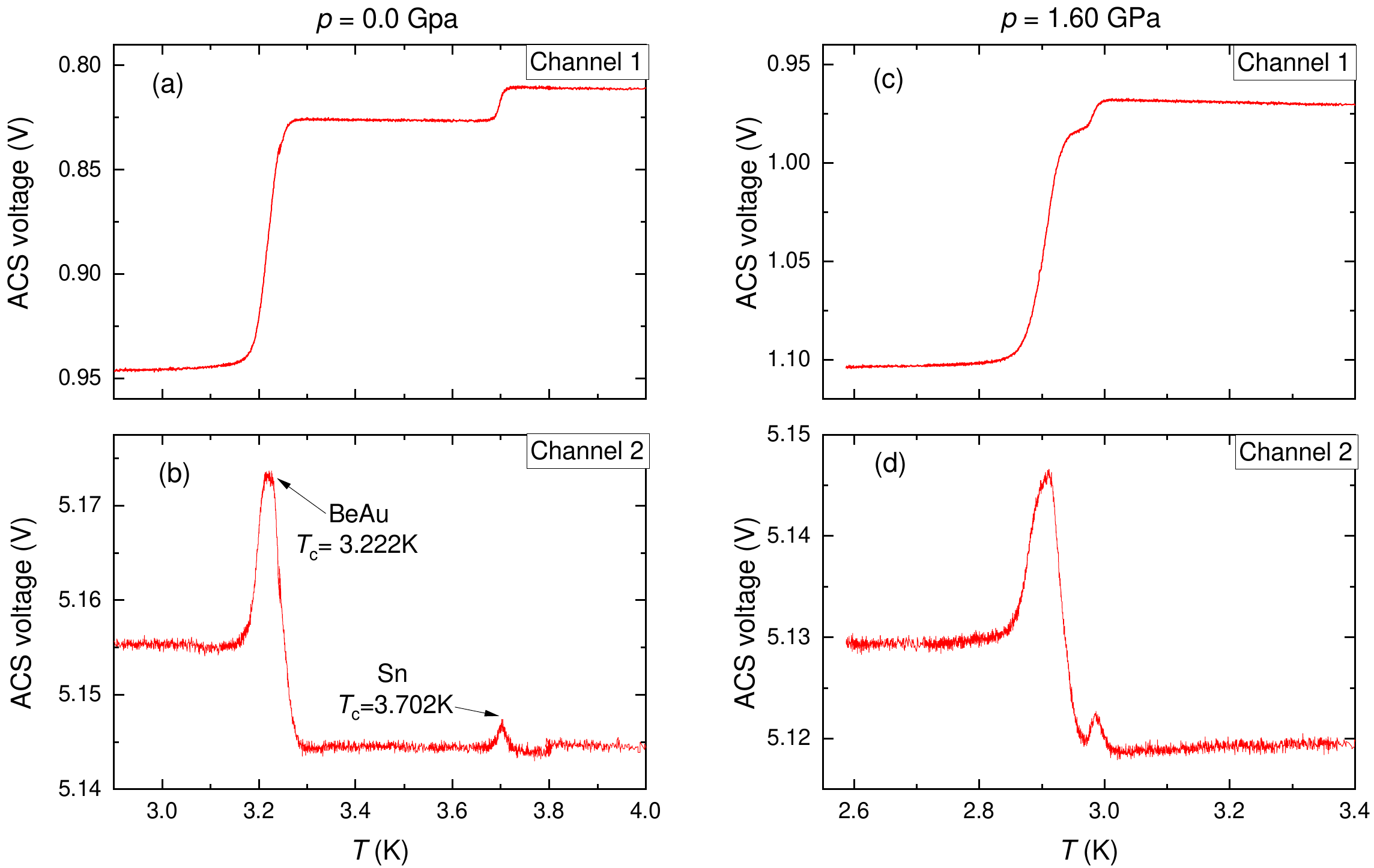}
\caption{(a) The AC susceptibility (ACS) response of BeAu sample and Sn pressure indicator measured at ambient pressure. The initial phase of the excitation signal was adjusted in order to measure the real part of AC susceptibility ($\chi\prime$) at the 'Channel 1' of the lock-in amplifier. (b) The ACS response of BeAu and Sn measured at the 'Channel 2' of the lock-in amplifier measured at $p=0$. The output of the 'Channel 2' is mostly determined by the imaginary part of AC susceptibility ($\chi\prime\prime$) with slight admixture of $\chi\prime$. (c) and (d) -- the same as in panels (a) and (b) but measured at $p\simeq 1.6$~GPa. }
 \label{fig:ACS}
\end{figure*}

\section{Equation of state}\label{seq:EOS}

The x-ray diffraction spectra were measured at $T \simeq 300$~K. The crystal structure is of the FeSi type with the space group $P2_13$. At ambient pressure ($p = 0.0$), the lattice constant $a = 4.6692(2)$~{\AA} was found to be in agreement with the literature data \cite{Amon_PRB_2018}. Figure~\ref{fig:Structure}~(a) shows the pressure dependence of the lattice constant $a$. With the pressure increasing from 0 to $\simeq 3$~GPa, the constant $a$ decreases by nearly 0.6\%.

The pressure dependence of the reduced volume [$V(p)/V_0$, where $V_0$ is the unit cell volume at $p = 0$] is shown in Fig.~\ref{fig:Structure}~(b). \textcolor{black}{By using the definitions from Refs.} \cite{Strassle_PRB_2014, Klotz_PRB_2017}:
\begin{equation}
 B_0=-\left(\frac{\partial p}{\partial \ln V}\right)_0, \ \
 B_0^\prime=-\left(\frac{\partial B_0}{\partial p}\right)_0, \nonumber
 \label{eq:EOS_parameters}
\end{equation}
($B_0$ and $B_0^\prime$ are the bulk modulus and their first derivative at ambient pressure) \textcolor{black}{the $V(p)/V_0$ data were fit using the Murnaghan equation of state (EOS) \cite{Murnaghan_AmJMath_1937}:}

\begin{equation}
p(V)=\frac{B_0}{B_0^\prime}\left[\left(\frac{V}{V_0}  \right)^{-B_0^\prime}-1\right]
 \label{eq:Murnaghan}
\end{equation}

The solid lines in Fig.~\ref{fig:Structure}~(b) correspond to the fit of \textcolor{black}{Murnaghan EOS} to the data. The fit parameters are $B_0=132.5(1.8)$~GPa and $B_0^\prime=29.6(1.9)$.

\section{Pressure dependence of the superconducting transition temperature $T_{\rm c}$}\label{sec:ACS}
\subsection{Experimental determination of $T_{\rm c}$ vs. $p$ behaviour}

\begin{figure}[htb]
\includegraphics[width=0.8\linewidth]{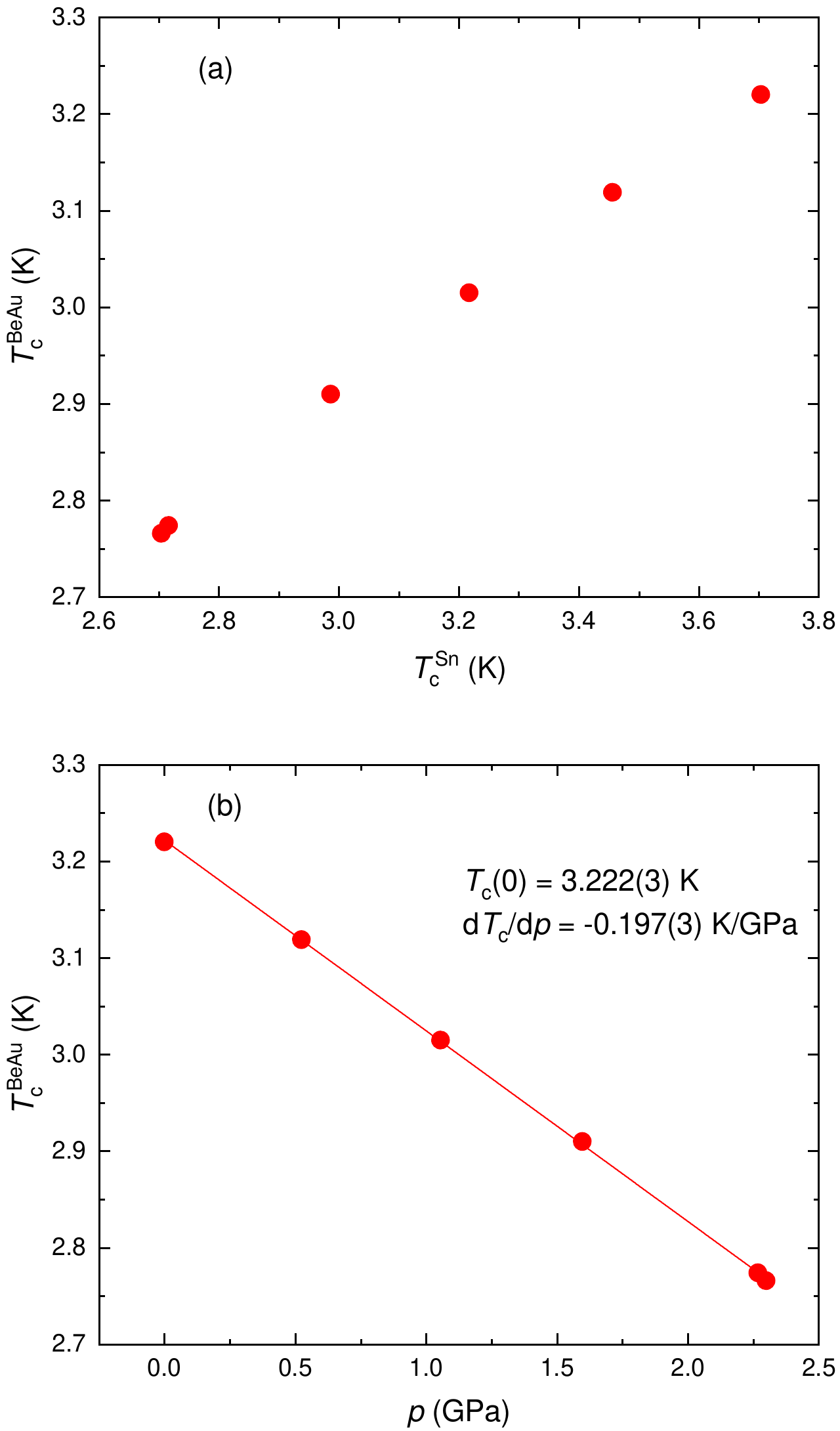}
\caption{(a) Dependence of $T_{\rm c}^{\rm BeAu}$ on $T_{\rm c}^{\rm Sn}$. (b) Dependence of $T_{\rm c}$ of BeAu on pressure. The linear fit (solid lines) result in the average slope ${\rm d} T_{\rm c}/{\rm d}p \simeq -0.197$~K/GPa.}
\label{fig:Tc_vs_p}
\end{figure}

The dependence of the superconducting temperature $T_{\rm c}$ on pressure was studied using the AC susceptibility (ACS) technique. Figure~\ref{fig:ACS} shows the ACS response measured at ambient pressure [panels (a) and (b)] and at $p \simeq 1.60$~GPa [panels (c) and (d)]. The outputs of 'Channel 1' and 'Channel 2' correspond mostly to the 'real' ($\chi'$, with the step-like change at the corresponding $T_{\rm c}$) and 'imaginary' ($\chi''$, with peaks at $T_{\rm c}$s) parts of the ACS responses, respectively. Note the small admixture of the $\chi'$ contribution to the BeAu response present in 'Channel 2'. The superconducting transition temperatures of BeAu ($T_{\rm c}^{\rm BeAu}$) and Sn ($T_{\rm c}^{\rm Sn}$) were determined \textcolor{black}{from the peak positions} as shown schematically in Fig.~\ref{fig:ACS}~(b). Figure~\ref{fig:Tc_vs_p}~(a) shows the dependence of $T_{\rm c}^{\rm BeAu}$ on $T_{\rm c}^{\rm Sn}$.

The pressure inside the cell was further determined from $T_{\rm c}^{\rm Sn}$, which, in accordance with the literature data, follows the relation \cite{Eiling_JPF_1981}:

\begin{equation}
T_{\rm c}^{\rm Sn}(p) = T_{\rm c}^{\rm Sn}(0) - 0.4823 \; p + 0.0207 \; p^2.
\label{eq:Tc_Sn}
\end{equation}

Here $T_{\rm c}^{\rm Sn}(0)$ is the superconducting transition temperature of Sn at $p=0$. The dependence of $T_{\rm c}^{\rm BeAu}$ on pressure is shown in Fig.~\ref{fig:Tc_vs_p}~(b). Solid line represent linear fits with the mean values of $T_{\rm c}^{\rm BeAu}(0) \simeq 3.222(3)$~K and the slope ${\rm d} T_{\rm c}/{\rm d}p \simeq -0.197(3)$~K/GPa, respectively.

The decrease in $T_{\rm c}$ of BeAu with increasing pressure is consistent with the behavior of most known superconductors (see, for example, Refs.~\onlinecite{Lorenz_2005, Schilling_book_2007, Schilling_JPCS_2008} and references therein). The value of the slope ${\rm d} T_{\rm c}/{\rm d}p \simeq -0.195$~K/GPa is approximately half of that for single-element conventional superconductors such as Sn ($\simeq -0.482$~K/GPa), In ($\simeq -0.381$~K/GPa), and Pb ($\simeq -0.365$~K/GPa) \cite{Eiling_JPF_1981}.

\subsection{Effect of pressure on the electron and phonon density of states}\label{sec:DFT-DOS}

\begin{figure}[htb]
	\centering
	\includegraphics[width=0.45\textwidth]{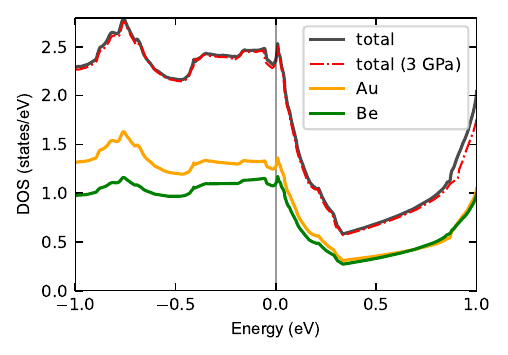}
	\caption{GGA DOS at ambient pressure (black) and 3 GPa (red, dash-dotted). Atom-resolved DOS of Be (Au) are green (gold). Fermi level is at
zero energy.}
	\label{fig:edos}
\end{figure}

In the experimentally measured pressure range up to 2.2~GPa, the superconducting critical temperature $T_{\rm c}$ of BeAu shows a slight decrease, almost linear in pressure. To rationalize this behavior, we first study how pressure affects the electronic structure of the metallic phase. The key quantity is the density of states (DOS) at the Fermi level, which controls the number of electrons that are able to form Cooper pairs; $T_{\rm c}$ is proportional to this quantity \cite{Bardeen_PR_1957}. The atom-resolved DOS (Fig.~\ref{fig:edos}) reveals comparable contributions from Be and Au at the Fermi level. Since superconductivity in BeAu is believed to be conventional \cite{Amon_PRB_2018}, we can expect that Be states dominate the electron-phonon coupling: due to its small atomic number, screening of the nuclear charge is small, allowing for a sizable electron-phonon interaction.

\begin{figure}[b!]
	\centering
	\includegraphics[width=0.45\textwidth]{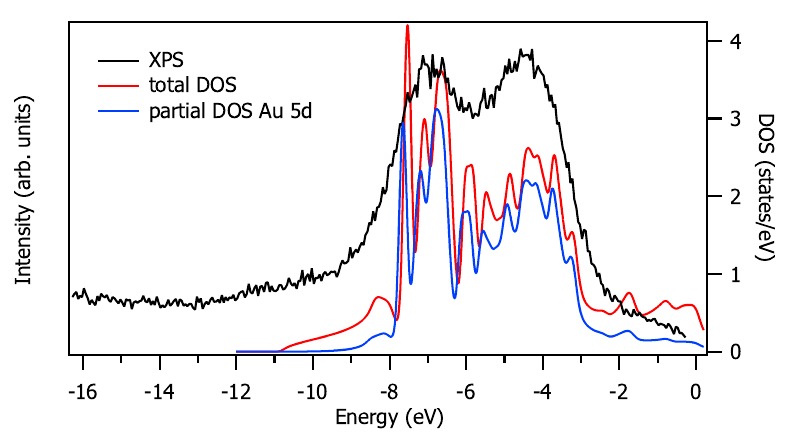}
	\caption{X-ray photoemission spectroscopy taken on BeAu, together with calculated density of states. \textcolor{black}{The broadening of the measured spectrum is due to the energy resolution determined by the high energy of incident light (1486.6 eV).}
}
	\label{fig:XPS}
\end{figure}

In order to support the DOS calculation by experimental data, we provide a measurement of the full valence band of BeAu by XPS in Fig.~\ref{fig:XPS}. The sample surface has been polished but no further cleaning attempt has been made prior to the measurement shown. At the photon energy of the measurement ($hv=1486.6$ eV) the photoemission intensity is dominated by the Au $5d$ orbitals. The agreement between experiment and theory is convincing, corroborating the above arguments regarding the low energy region.

\textcolor{black}{Minute differences between the DOS at ambient pressure and at 3\,GPa (Fig.~\ref{fig:edos}) indicate that the pressure-induced decrease of $T_{\rm c}$ is not of a purely electronic origin. Instead, it should be attributed to changes in the vibrational properties of BeAu.} For conventional phonon-mediated superconductivity, the Migdal-Eliashberg theory \cite{EliashbergReview} generally suggests that high-frequency modes contribute less to the effective electron-electron interaction. Since external pressure typically gives rise to a stiffening of phonons -- similar to hardening the spring constant in simple spring-chain model -- it generally lowers the $T_{\rm c}$. In BeAu, due to the sizable difference in atomic mass of the constituent elements, we expect predominant changes in the vibrational modes to stem from Be. The corresponding phonon modes are expected to have a stronger impact on the superconducting pairing.

\begin{figure}[htb]
	\centering
	\includegraphics[width=0.45\textwidth]{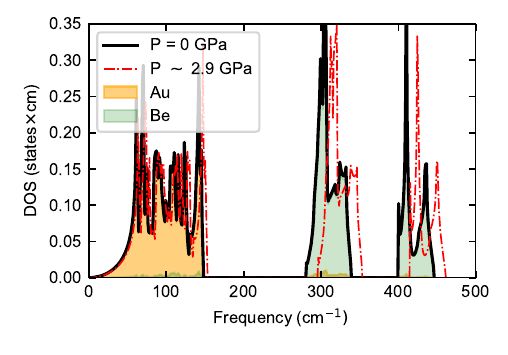}
	\caption{Phonon DOS at ambient pressure and 3 GPa, in green (gold) we highlight the partial contribution of Be (Au). Note that pressure shifts Be-dominated optical modes to higher frequencies, leaving the Au-dominated modes intact.}
	\label{fig:phdos}
\end{figure}

To verify this scenario, we performed harmonic phonon calculations. In the phonon DOS (Fig.~\ref{fig:phdos}), we observe a remarkable separation between the low-energy acoustic modes dominated by Au, and the optical modes contributed by Be.  Hence, in contrast to the electronic structure, where Au and Be states are largely hybridized, oscillations of Be and Au are nearly independent.  Another significant consequence of the mass difference between Au and Be is the behavior of the phonon modes under pressure: Au modes are essentially unaffected by strain, while Be modes are shifted to higher energies.

The remarkable separation of Au and Be vibrational modes allows us to construct a simplified model and compute how the phonon shift affects the critical temperature. Our starting point is the Allen and Dynes equation \cite{AllenDynes}:

\begin{equation}
        T_{\rm c} = \frac {\omega_{log}} {1.12} \exp \left[
	\frac
        {1.04(1+\lambda)}
        {\lambda-\mu^*(1+0.62\lambda)}
        \right],
    \label{eq:tc}
\end{equation}

\noindent where $\lambda$ is a \textcolor{black}{dimensionless} measure of the strength of the electron-phonon propagator and $\omega_{log}$ is a cut-off frequency that replaces the Debye cut-off in the BCS model. They both depend on the electron-phonon propagator, as described by the Eliashberg theory. The parameter $\mu^*$ quantifies the strength of electrostatic repulsion, and it is generally taken to be 0.11 for most superconductors \cite{ReviewH}.

\begin{figure}[htb]
	\centering
	\includegraphics[width=0.45\textwidth]{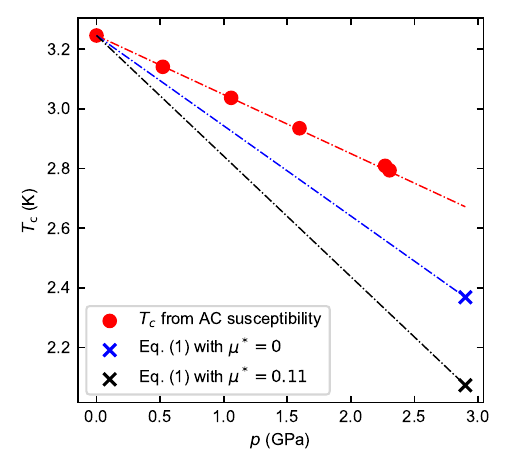}
	\caption{Pressure dependence of the critical temperature $T_{\rm c}$. Lines are guide to the eye.}
	\label{fig:eph}
\end{figure}

Under the assumptions that i) Au phonons do not contribute to the superconducting pairing, and ii) the electron-phonon coupling constant does not change under pressure, we can make explicit the dependency of both $\lambda$ and $\omega_{log}$ on the phonon DOS. In particular, we find $\lambda$ to be given by the equation:
\begin{equation}
	\lambda = \gamma \int_{\omega_1}^{\infty} d\omega \frac {\rho(\omega)}
	{\omega^2},
	\label{eq:eph}
\end{equation}

\noindent where $\rho(\omega)$ is the phonon DOS, and the proportionality constant $\gamma$ includes the electronic part of the electron-phonon interaction. A similar relation is also derived for $\omega_{log}$, which is, however, dependent only on the phonon properties.  Evaluation of both expressions is detailed in the Appendix 2. The proportionality constant $\gamma$ is the only quantity which is not computed; we extract $\gamma$ from the measured values of $T_{\rm c}$ at 0 GPa and estimate the effect of the phonon shift through Eq.~\ref{eq:eph}. The results for $\mu^* = 0$ and $\mu^* = 0.11$ are summarized in Fig.~\ref{fig:eph}.
As expected, we find that $T_{\rm c}$ decreases with pressure. With the enhanced Coulomb repulsion ($\mu^* = 0.11$) $T_{\rm c}$ drops more rapidly, but the variations are not drastic. Moreover, for $\mu^* = 0.11$ and at 0 GPa, we evaluate $\lambda = 0.45$, which aligns with the experimentally reported value of 0.5 in Ref. \cite{Amon_PRB_2018}. The good agreement with experimental data \emph{a posteriori} justifies our assumption of negligible Au contribution to the electron-phonon coupling and corroborates our qualitative analysis based on the phonon DOS.

To summarize, the applied pressure induces a stiffening of the phonon modes, mainly associated with Be, ultimately lowering the $T_{\rm c}$ in line with the Migdal-Eliashberg theory. Hence, superconductivity in BeAu is similar to that in binary and ternary hydrogen-based superconductors, where the mass difference is further accentuated -- with hydrogen being the lightest element.

\section{Pressure dependence of the thermodynamic critical field $B_{\rm c}$}\label{sec:Bc}

Measurements of the pressure dependence of the thermodynamic critical field $B_{\rm c}$ were performed by means of $\mu$SR technique. The experiments were conducted in the so-called intermediate state, {\it i.e.}, when the magnetic field penetrates type-I superconducting materials in the form of normal state domains. The field range for domain formation is determined by the demagnetization factor $N$, and it occurs for applied fields ($B_{\rm ap}$) ranging in 
\begin{equation}
(1-N)B_{\rm c} \leq B_{\rm ap} \leq B_{\rm c}
 \label{eq:intermediate-state}
\end{equation} 
(see, for example, Refs.~\onlinecite{Huebener_book_2001, Karl_PRB_2019} and references therein).

\textcolor{black}{In TF-$\mu$SR experiments, the formation of the intermediate state is clearly visible through the appearance of a well-pronounced 'double-peak' structure. The peak at zero-field corresponds to the response of the Meissner state (superconducting, SC) domains, which completely expel the magnetic field from their volumes ($B_{\rm SC}=0$). The second peak is determined by the response of the nonsuperconducting (normal state, NS) domains, where superconductivity is destroyed due to the fact that the field inside NS domains becomes exactly equal to the thermodynamic critical field ($B_{\rm NS}=B_{\rm c}$). The intensities of each peak are proportional to the corresponding normal-state and superconducting-state volume fractions. It is important to note that the intermediate state is formed with an applied field smaller than $B_{\rm c}$ ($B_{\rm ap}<B_{\rm c}$, see Eq.~\ref{eq:intermediate-state}), meaning that the NS peak always stays above the applied field (see, {\it e.g.}, Refs.~\onlinecite{Karl_PRB_2019, Khasanov_BiII_PRB_2019} and references therein). }

In the experiments described below, each measured point was reached in two steps: first by stabilizing the temperature and second by sweeping the field to $B > B_{\rm c}$ (up to 30~mT in our case) and then decreasing it back to the desired value. At each particular temperature, two measurements were conducted at $B_{\rm ap} \simeq 0.6 B_{\rm c}(T)$ and $\simeq 0.3 B_{\rm c}(T)$).

\subsection{Determination of the pressure evolution of the thermodynamic critical field  by means of TF-$\mu$SR}

\begin{figure}[htb]
\includegraphics[width=0.8\linewidth]{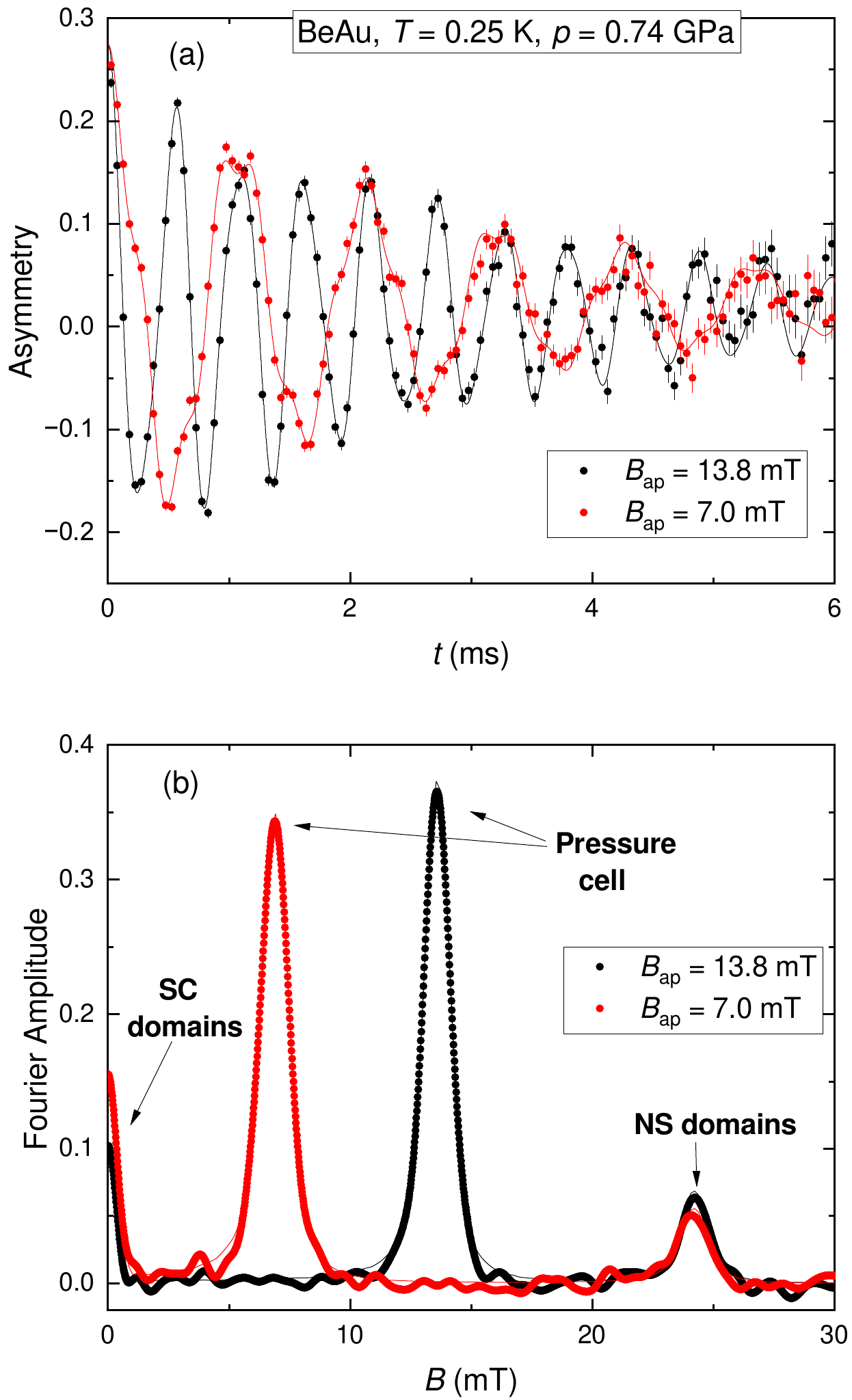}
%
\caption{(a) TF-$\mu$SR time spectra taken at applied magnetic fields $B_{\rm ap}=13.8$ and $7.0$~mT, $T\simeq 0.25$~K and $p\simeq 0.74$~GPa. Solid lines are fits by means of Eq.~\ref{eq:P(t)} to the data (see text for further details). (b) The Fourier transform of TF-$\mu$SR time spectra presented in (a).}
 \label{fig:Time-spectra_full}
\end{figure}

Figure \ref{fig:Time-spectra_full}~(a) shows the TF-$\mu$SR time spectra taken at applied magnetic fields $B_{\rm ap}=13.8$ and $7.0$~mT, $T\simeq 0.25$~K, and $p\simeq 0.74$~GPa. The corresponding Fourier transforms are shown in Fig.~\ref{fig:Time-spectra_full}~(b). The magnetic field distribution is typical for a type-I superconductor in the intermediate state. Three pronounced peaks originating from the muons stopped in the Meissner state domains ($B\equiv 0$), normal state domains ($B\equiv B_{\rm c}\simeq 24$~mT), and in the pressure cell walls ($B=B_{\rm ap}$) \cite{Karl_PRB_2019, Khasanov_BiII_PRB_2019}, are clearly visible. Note that up to the highest pressures reached in our experiments ($\simeq 2.2$~GPa), the response of BeAu always corresponds to type-I superconducting behavior.

\begin{figure*}[htb]
\includegraphics[width=1.0\linewidth]{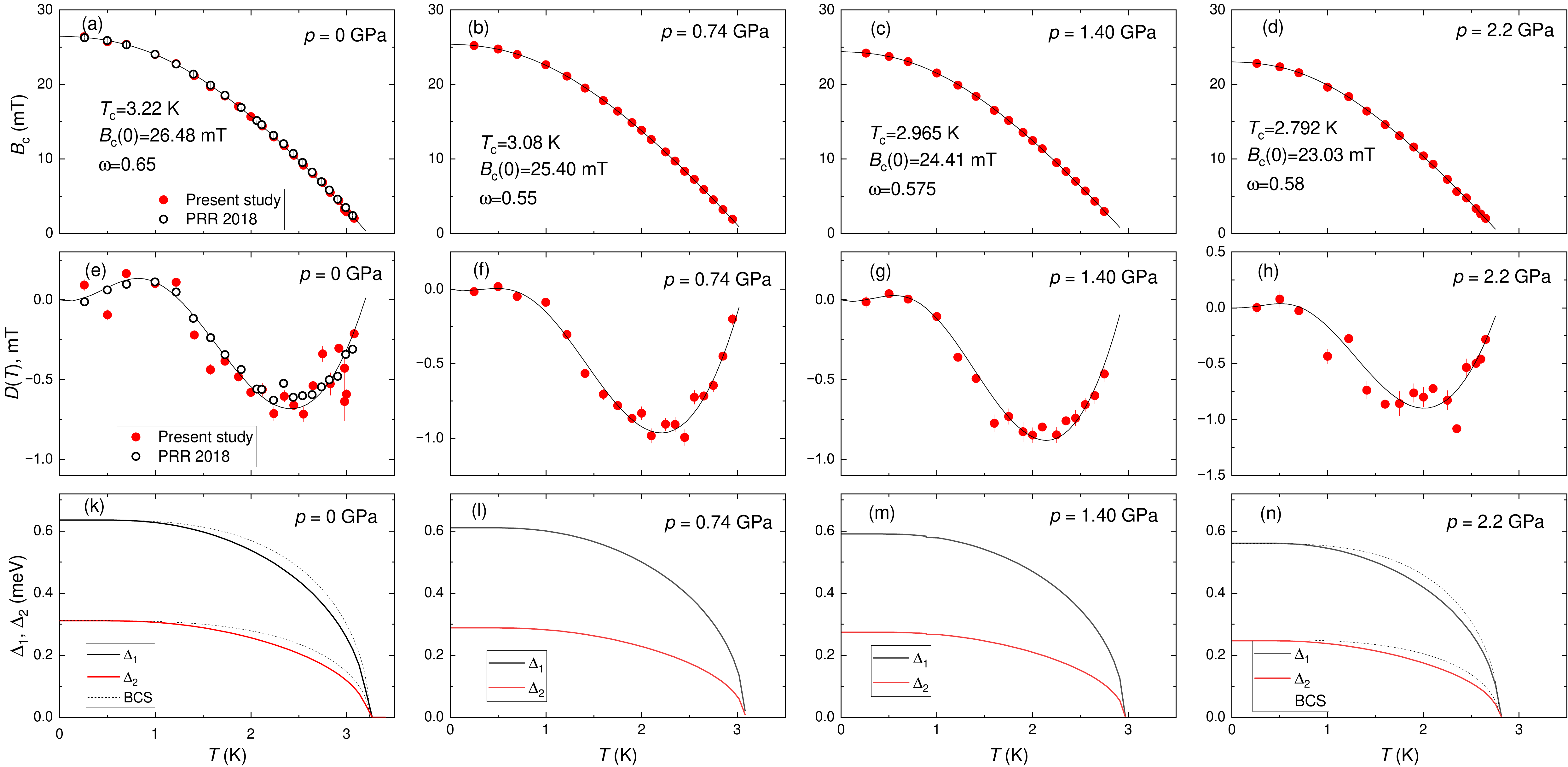}
\caption{(a)-(d) Dependence of the thermodynamic critical field $B_{\rm c}$ on temperature measured at pressures $p=0.0$, 0.74, 1.4, and 2.2~GPa. (e)-(h) The deviation function $D(T)=B_{\rm c}(T)-B_{\rm c}(0)[1-(T/T_{\rm c})^2]$. The solid lines in panels (a)-(h) are fits within the framework of self-consistent two-gap model. (k)-(m) Temperature evolution of the big ($\Delta_1$) and the small ($\Delta_2$) superconducting energy gaps obtained from the fit of self-consistent two-gap model to $D(T)$ data. The dashed lines in (k) and (n) represent temperature dependence of a BCS gap.}
 \label{fig:Bc_vs_T}
\end{figure*}

The solid lines in Fig.~\ref{fig:Time-spectra_full}~(a) correspond to the fit of the following functional form to the data \cite{Khasanov_PRL_2010, Khasanov_PRB_2011}:
\begin{equation}
A_0 P(t) = A_{\rm pc} P_{\rm pc}(t) + A_{\rm s} P_{\rm s}(t) \label{eq:P(t)}
\end{equation}
\noindent here, $A_0$ is the initial asymmetry of the muon-spin ensemble. Indexes denote the sample (s) and the pressure cell (pc) contributions. $A_{\rm pc}$ ($A_{\rm s}$) and $P_{\rm pc}(t)$ [$P_{\rm s}(t)$] are the asymmetry and the time evolution of the muon spin polarization for muons stopped in the pressure cell and the sample. The pressure cell contribution is given by:
\begin{equation}
P_{\rm pc}(t) = e^{-\lambda_{\rm pc} t} \cos (\gamma_\mu B_{\rm ap} t + \phi),
\end{equation}
\noindent where $\lambda_{\rm pc}$ is the exponential relaxation rate, $\phi$ is the initial phase of the muon-spin ensemble, and $\gamma_\mu = 2\pi \cdot 135.5342$~MHz/T is the muon gyromagnetic ratio.

The sample contribution is described by assuming a separation between the normal state (NS) and the superconducting (SC) domains:
\begin{eqnarray}
P_{\rm s}(t) &=& f_{\rm NS} e^{-\lambda_{\rm NS} t} \cos (\gamma_\mu B_{\rm NS} t + \phi) + \nonumber\\
&& (1 - f_{\rm NS}) \left[ \frac{1}{3} + \frac{2}{3}(1 - \sigma_{\rm GKT}^2 t^2) e^{-\sigma_{\rm GKT}^2 t^2 / 2} \right]. \label{eq:Sample_P(t)}
\end{eqnarray}
\noindent where the first term on the right-hand side of Eq.~\ref{eq:Sample_P(t)} corresponds to the sample's normal state response: $f_{\rm NS}$ is the normal state volume fraction ($f_{\rm NS}=1$ for $T\geq T_{\rm c}$), $\lambda_{\rm NS}$ is the exponential relaxation rate, and $B_{\rm NS}$ is the internal field ($B_{\rm NS}=B_{\rm c}$ for $T<T_{\rm c}$ and $B_{\rm NS}=B_{\rm ap}$ for $T\geq T_{\rm c}$). The second term describes the contribution of the superconducting part of the sample remaining in the Meissner state ($B_{\rm SC}=0$). It is approximated by the Gaussian Kubo-Toyabe function with the relaxation rate $\sigma_{\rm GKT}$, which is generally used to describe the nuclear magnetic moment contribution in zero-field $\mu$SR experiments (see Ref.~\onlinecite{Yaouanc_book_2011} and references therein).

Figure~\ref{fig:Bc_vs_T}~(a)-(d) shows the temperature dependencies of the thermodynamic critical field $B_{\rm c}$ of BeAu measured at $p=0.0$, 0.74, 1.4, and 2.2~GPa. Three important points need to be considered:
\begin{enumerate}
 \item Pressure does not change the nature of BeAu superconductivity. Up to the highest pressure reached in the experiment ($\simeq2.2$~GPa), BeAu remains a type-I superconductor.
 \item The absolute values of the critical field and the superconducting transition temperatures decrease with increasing pressure. The zero-temperature value of the critical field $B_{\rm c}(0)$ decreases from $B_{\rm c}(0) \simeq 26.5$~mT at $p=0.0$ to $\simeq 23$~mT at $p \simeq 2.2$~GPa, while $T_{\rm c}$ decreases by $\sim0.45$~K (from $3.22$~K down to $\simeq 2.79$~K) within the same pressure range. The decrease of $T_{\rm c}$ obtained in $\mu$SR studies agrees with that obtained in ACS experiments (see Sec.~\ref{sec:ACS}).
 \item In panel (a), we have also added the ambient pressure $B_{\rm c}(T)$ data from Ref.~\onlinecite{Khasanov_AuBe_PRR_2020}. Note the one-to-one agreement between the ambient pressure $B_{\rm c}(T)$ data measured without a pressure cell at a low-background Dolly spectrometer in Ref.~\onlinecite{Khasanov_AuBe_PRR_2020} and the $B_{\rm c}(T)$ data obtained in the present study.
\end{enumerate}

\subsection{Self-consistent two-gap fits to $B_{\rm c}(T)$ data}

\begin{figure}[htb]
\includegraphics[width=0.8\linewidth]{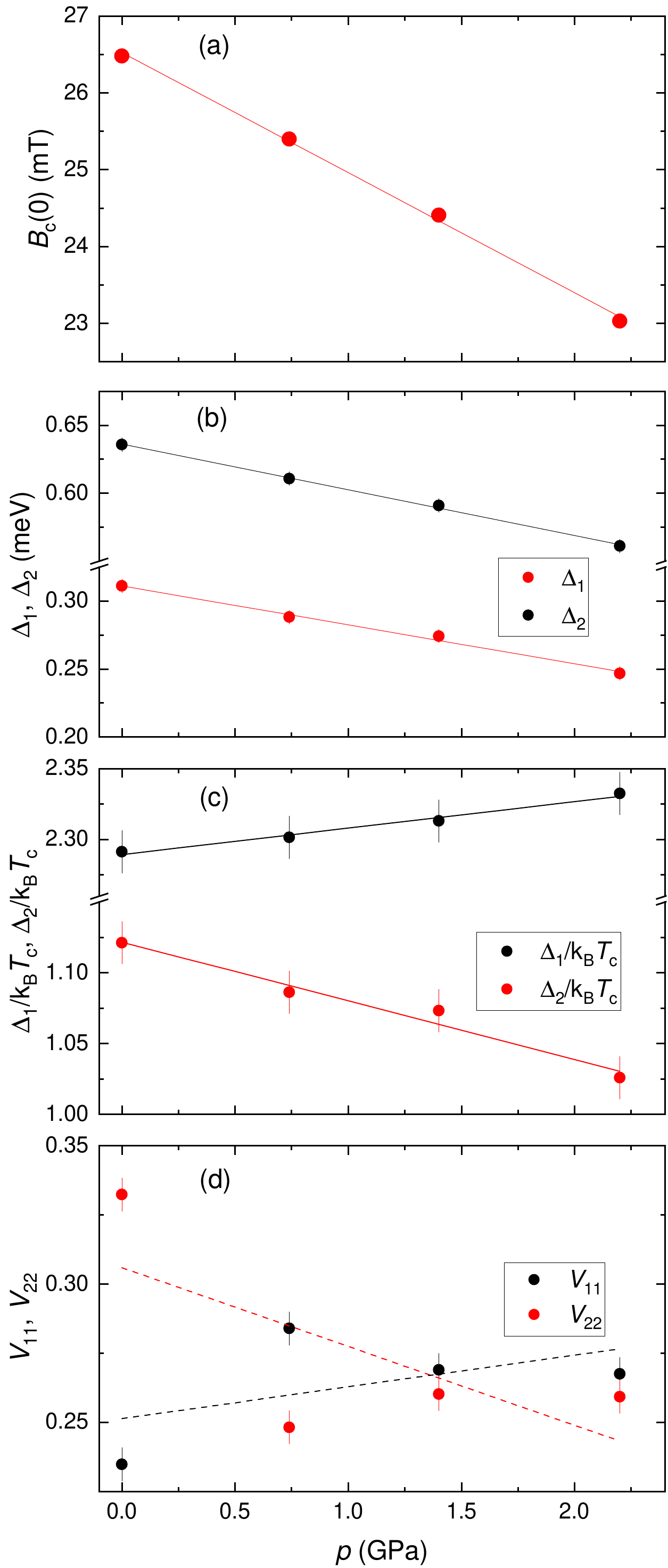}
%
\caption{(a) Pressure dependence of the zero-temperature value of the thermodynamic critical field $B_{\rm c}(0)$. \textcolor{black}{(b)  Dependencies of the big ($\Delta_1$) and the small ($\Delta_2$) superconducting energy gaps on pressure}. (c) Dependence of $\Delta_1/k_{\rm B}T_{\rm c}$ and $\Delta_1/k_{\rm B}T_{\rm c}$ on $p$. (d) The evolution of the intraband potentials $V_{11}$ and $V_{22}$ with pressure. The solid lines are linear fits to the data.}
 \label{fig:Bc_fit_parameters}
\end{figure}

At ambient pressure, BeAu is known to be a two-gap superconductor with distinct gap-to-$T_{\rm c}$ ratios. This was confirmed through the analysis of specific heat and $B_{\rm c}(T)$ data \cite{Khasanov_AuBe_PRR_2020, Khasanov_AuBe_PRB_2020}, as well as by recent tunneling experiments \cite{Datta_PRB_2022}. All of these previous studies agree with our value $\Delta_1/k_{\rm B}T_{\rm c} \simeq 2.2$ and $\Delta_2/k_{\rm B}T_{\rm c} \simeq 1.2$ for the big ($\Delta_1$) and the small ($\Delta_2$) superconducting energy gaps, as extracted from Fig.\ \ref{fig:Bc_fit_parameters}.

The deviations of the $B_{\rm c}$ {\it vs.} $T$ curves from the parabolic function: $D(T)=B_{\rm c}(T)-B_{\rm c}(0)[1-(T/T_{\rm c})^2]$ are presented in Figs.~\ref{fig:Bc_vs_T}~(e)-(h). Following Refs.~\onlinecite{Padamsee_JLTP_1973, Johnston_SST_2013}, the shape of $D(T)$ depends strongly on the $2\Delta/k_{\rm B}T_{\rm c}$ ratio and is also sensitive to the symmetries and the magnitudes of the superconducting energy gaps. Considering the multiple-gap behavior of BeAu at ambient pressure, the $D(T)$ data were analyzed within the self-consistent two-gap approach as described in Refs.~\onlinecite{Khasanov_AuBe_PRR_2020, Khasanov_AuBe_PRB_2020} (see also Sec.~\ref{sec:self-consistent_model}). The total number of fit parameters is six, which includes four coupling parameters [two intraband ($V_{11}$ and $V_{22}$), and two interband ($V_{12}$ and $V_{21}$) interaction potentials], the zero-temperature value of the thermodynamic critical field $B_{\rm c}(0)$, and the contribution of the larger gap to the free energy $\omega$. The ambient pressure value of the Debye frequency $\omega_{\rm D}=33.4$~meV was taken from Ref~\onlinecite{Amon_PRB_2018}. The increase of $\omega_{\rm D}$ with increasing pressure was assumed to follow:
\begin{equation}
  \omega_{\rm D}(p)=\omega_{\rm D}(0)\Big(1+\frac{\gamma_{\rm G}~p}{B_0}\Big),
  \label{eq:omega_D}
\end{equation}
which is a consequence of the Gr\"{u}neisen equation $\gamma_{\rm G}=-{\rm d}\ln \omega_{\rm D}/{\rm d}\ln V$ [$\gamma_{\rm G}$ is the Gr\"{u}neisen parameter, $B_0 \simeq 131$~GPa is the bulk modulus (see Sec.~\ref{seq:EOS}), and $V$ is the sample volume]. The Gr\"{u}neisen parameter was assumed to be $\gamma_{\rm G}=3.0$, which is close to the corresponding value for Au and is more than three times as large compared to Be. We should, however, emphasize that the parameters of the self-consistent model are not very sensitive to the exact value of $\gamma_{\rm G}$. For example, increasing the Gr\"{u}neisen parameter $\gamma_{\rm G}$ from 0.0 to 3.0 leads to a corresponding increase of $\omega_{\rm D}$ by $\sim$ 3.3\% and a decrease of the coupling strength parameters by $\sim$ 0.4\% at $p=2.5$~GPa.

The fit of the self-consistent model to the data requires solving two coupled nonlinear gap equations (Eq.~\ref{eq:Self-consistent_model}) to obtain the temperature evolutions of the larger $\Delta_1(T)$ and smaller $\Delta_2(T)$ superconducting energy gaps. $\Delta_1(T)$ and $\Delta_2(T)$ are further used to derive the $B_{\rm c}(T)$ curve by solving the free energy equations described by Eqs.~\ref{eq:Critical-field_two-gap}, \ref{eq:Fn}, and \ref{eq:Fs}. We used PTC Mathcad Express with the Levenberg-Marquardt nonlinear equation solver \cite{Mathcad}. The results of the analysis are shown by black solid lines in panels (a)-(d) and (e)-(h) of Fig.~\ref{fig:Bc_vs_T} for $B_{\rm c}(T)$ and $D(T)$, respectively. The temperature dependencies of the larger and smaller superconducting gaps are shown in Figs.~\ref{fig:Bc_vs_T}~(k)-(m). The values of the fit parameters are summarized in Table~\ref{tab:table2}, and several of them are plotted in Fig.~\ref{fig:Bc_fit_parameters}.

\begin{table}[htb]
  \begin{center}
    \caption{The results of the various EOS fits: $B_0$ is the bulk modulus at ambient pressure and $B_0^\prime$ the first derivative of $B_0$at ambient pressure.}
    \label{tab:table2}
    \begin{tabular}{l|cccccccc}
      $p$  &$B_{\rm c}(0)$ &$T_{\rm c}$&$\Delta_1$&$\Delta_2$&$V_{11}$&$V_{22}$&$V_{12}$&$V_{21}$\\
      (GPa)&(mT)           &(K)        &(meV)     &(meV)     &        &        &        &        \\
      \hline
      0.0  &26.48&3.22&0.64&0.31&0.235&0.332&0.312&0.0636\\
      0.74 &25.40&3.08&0.61&0.29&0.284&0.248&0.233&0.0699\\
      1.40 &24.41&2.97&0.59&0.27&0.269&0.260&0.244&0.0662\\
      2.20 &23.03&2.79&0.56&0.25&0.268&0.259&0.243&0.0561\\
    \end{tabular}
  \end{center}
\end{table}

From the analysis of $B_{\rm c}(T)$ of BeAu within the two-gap approach, the following important points follow:
\begin{enumerate}
  \item The zero-temperature value of the thermodynamic critical field decreases with increasing pressure [see Fig.~\ref{fig:Bc_fit_parameters}~(a) and Table~\ref{tab:table2}]. The linear fit results in ${\rm d}B_{\rm c}(0)/{\rm d}p=-2.65(1)$~mT/GPa.
  \item The temperature dependencies of the big ($\Delta_1$) and the small ($\Delta_2$) energy gaps are weaker than the BCS prediction [see, {\it e.g.}, Figs.~\ref{fig:Bc_vs_T}~(k) and (n)]. This partly agrees with the results of tunneling experiments by Datta {\it et al.}\cite{Datta_PRB_2022}, which reported weaker than BCS behavior for the smaller gap. Note that for a `classical' two-gap superconductor MgB$_2$, the deviation of both gaps from BCS-type temperature behavior was reported \cite{Gonnelli_PRL_2002, Iavarone_PRL_2002, Chen_APL_2008, Kim_Symmetry_2019}.
  \item \textcolor{black}{The zero-temperature values of both, $\Delta_1$ and $\Delta_2$, energy gaps decrease linearly with increasing pressure [Fig.~\ref{fig:Bc_fit_parameters}~(b)]. This behaviour stays inline with the pressure induced decrease of the superconducting transition temperature [Fig.~\ref{fig:Tc_vs_p}~(b)] and the thermodynamic critical field $B_{\rm c}$ [Fig.~\ref{fig:Bc_fit_parameters}~(a)]. The linear fits results in ${\rm d}\Delta_1/{\rm d}p=-0.0336(1)$~meV/GPa and ${\rm d}\Delta_2/{\rm d}p=-0.0286(2)$~meV/GPa for the big and the small gap, respectively. }
  \item In superconductors the gap-to-$T_{\rm c}$ ratio (often referred to as the 'coupling strength'), is typically compared to the weak-coupling BCS prediction $\Delta^{\rm BCS}/k_{\rm B}T_{\rm c}\simeq1.76$ \cite{Tinkham_book_1975}. Consequently, superconductors with a $\Delta/k_{\rm B}T_{\rm c}$ ratio comparable to or smaller than 1.76 are called 'weakly-coupled', while those with ratios exceeding 1.76 are termed 'strongly-coupled'. In BeAu, $\Delta_1/k_{\rm B}T_{\rm c} \sim 2.3$ and $\Delta_2/k_{\rm B}T_{\rm c} \sim 1.1$ [see Fig.~\ref{fig:Bc_fit_parameters}~(b)], suggesting strong coupling of the supercarriers within the bands where the 'large' superconducting gap opens and weak coupling within the bands with the 'small' superconducting energy gap.
  \item Figure~\ref{fig:Bc_fit_parameters}~(b) implies that $\Delta_1/k_{\rm B}T_{\rm c}$ increases, while $\Delta_2/k_{\rm B}T_{\rm c}$ decreases with increasing pressure. This suggests that pressure enhances (weakens) the coupling strength between the supercarriers within the bands where the large (small) superconducting energy gap has opened. This statement is further confirmed by the corresponding increase/decrease of the intraband potentials $V_{11}$ and $V_{22}$ [see Fig.~\ref{fig:Bc_fit_parameters}~(c) and Table~\ref{tab:table2}].
  \item The interband coupling potentials $V_{12}$ and $V_{21}$ have relatively high values. In particular, $V_{12}$ is comparable to $V_{11}$ and $V_{22}$ (see Table~\ref{tab:table2}). This suggests strong coupling between the bands, resulting in almost identical temperature dependencies of the larger [$\Delta_{1}(T)$] and smaller [$\Delta_{2}(T)$] gaps, as shown in Figs.~\ref{fig:Bc_vs_T}~(k)-(n). Note that in the case of weak interband coupling {\it i.e.} for $V_{12}, V_{21} \ll V_{11}, V_{22}$, the temperature behavior of $\Delta_{1}(T)$ and $\Delta_{2}(T)$ are very different \cite{Kogan_PRB_2009, Khasanov_PRL_2010, Gupta_FrontPhys_2019}.
\end{enumerate}

\section{The scaling relation between $B_{\rm c}(0)$ and $T_{\rm c}$}\label{sec:Bc_vs_Tc}

For conventional phonon-mediated superconductors, microscopic theories such as BCS \cite{Bardeen_PR_1957}, or the more general Eliashberg theory \cite{Eliashberg_1960a,Eliashberg_1960b, Carbotte_RMP_1990, Marsiglio_book_2008, Scalapino_book_1969, McMillan_book_1969, Allen_book_1982}, which account for retardation effects, were very successful in describing electronic properties. In particular, it was found that an empirical scaling relation between the zero-temperature value of the thermodynamic critical field $B_{\rm c}(0)$ and the transition temperature $T_{\rm c}$ exists for conventional type-I superconducting materials \cite{Rohlf_Book_1994, Khasanov_Ga_PRB_2020}.

\begin{figure}[htb]
\includegraphics[width=1\linewidth]{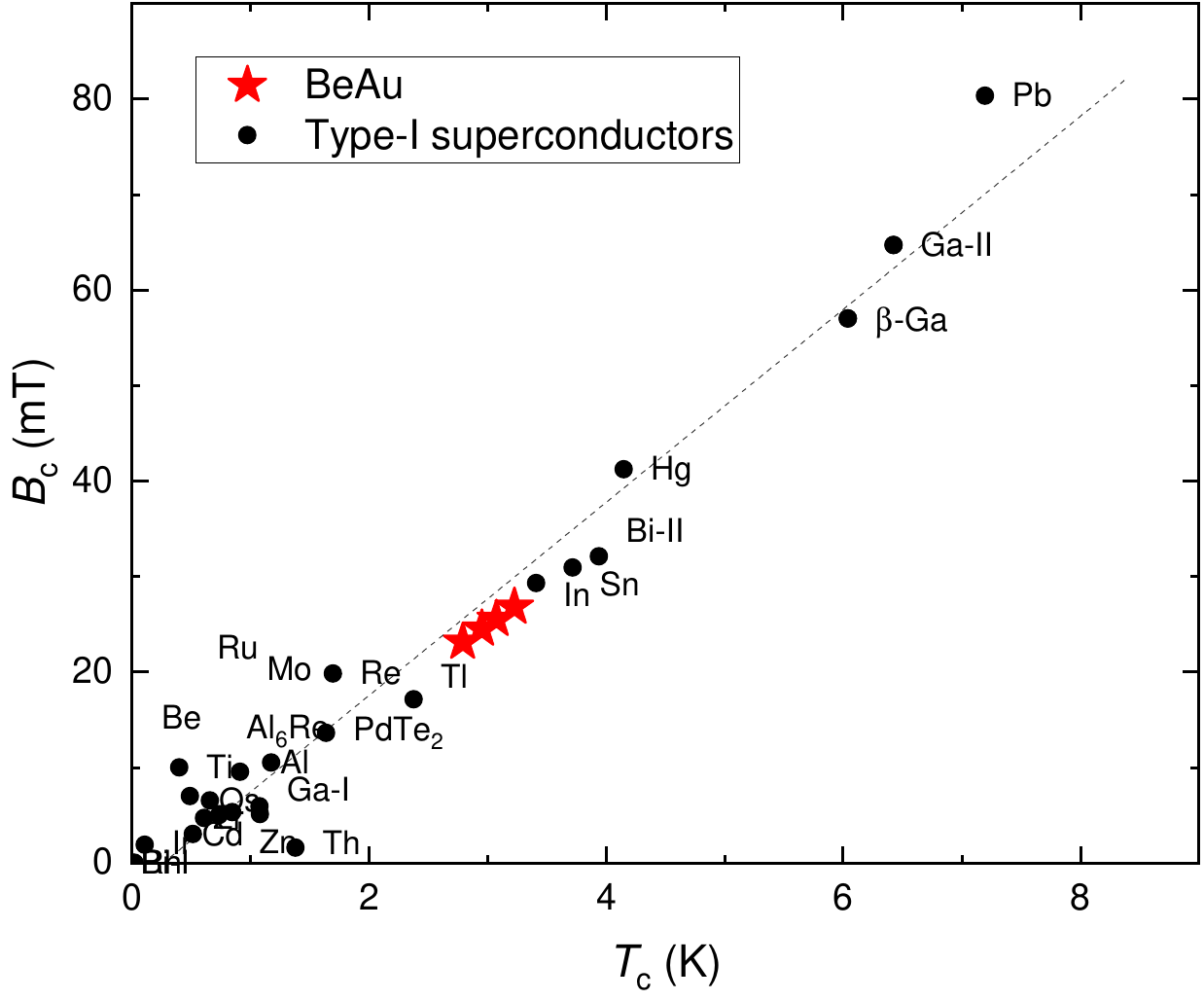}
%
\caption{Empirical relation between the zero-temperature value of the thermodynamic critical field $B_{\rm c}(0)$ and the transition temperature $T_{\rm c}$ for type-I superconductors, after Refs. \cite{Rohlf_Book_1994, Khasanov_Ga_PRB_2020}. Black symbols are data points from Refs. \cite{Kittel_Book_1996, Poole_Book_2014, Leng_PRB_2017, Prakash_Science_2017, Campanini_PRB_2018, Khasanov_BiII_PRB_2019, Beare_PRB_2019, Khasanov_Al_PRB_2021, Khasanov_Pb_PRB_2021, Arushi_PRB_2021, Leng_PRB_2019}, while the red stars represent BeAu from the current study.}
 \label{fig:Bc_vs_Tc}
\end{figure}

As established in Refs.~\onlinecite{Khasanov_AuBe_PRR_2020, Khasanov_AuBe_PRB_2020, Datta_PRB_2022} at ambient pressure, as well as confirmed in the above-reported high-pressure experiments, BeAu is a type-I superconductor. Unlike other conventional type-I superconducting materials, BeAu has two energy gaps with distinct gap-to-$T_{\rm c}$ ratios. It is important, therefore, to check if the $B_{\rm c}(0)$ {\it vs.} $T_{\rm c}$ scaling relation holds for BeAu.
Figure~\ref{fig:Bc_vs_Tc} shows the dependence of $B_{\rm c}(0)$ on $T_{\rm c}$ for various type-I superconductors. The values from the present study are denoted by red stars. Clearly, the obtained $B_{\rm c}(0)$ {\it vs.} $T_{\rm c}$ data for BeAu follow the general trend established for conventional type-I superconductors.

\section{Conclusions}\label{sec:conclusions}
Our results on BeAu under hydrostatic pressure reveal several key findings:

\begin{enumerate}
\item Pressure reduces the transition temperature $T_{\rm c}$.
\item The type of superconductivity remains type-I, unaffected by the applied pressure.
\item The zero-temperature value of the critical magnetic field $B_{\rm c}$ follows $T_{\rm c}$, and the observed values are consistent with the universal $B_{\rm c}$ versus $T_{\rm c}$ behavior established for phonon-mediated type-I superconductors.
\item The Au contribution to the electron-phonon coupling is negligible.
\item Application pressure induces a stiffening of the phonon modes, reminiscent of hydrogen-based superconductors.
\end{enumerate}

Overall, BeAu is one of a few noncentrosymmetric type-I superconductors known to date. Since its crystal structure is rather simple and it does not host substantial electron correlations, it is also a convenient material to study with theoretical means. Superconductivity of BeAu remains to be of interest due to its chiral B20 structure. The biggest obstacle in further exploring peculiar superconductivity of BeAu lies in the lack of single crystals. And even though this task is certainly challenging, overcoming it is certain to expand our understanding of superconductivity not only in BeAu, but also in other systems.

\appendix

\section{The self-consistent two-gap model of thermodynamic critical field}\label{sec:self-consistent_model}

Following Refs.~\cite{Bussmann-Holder_EPB_2004, Bussmann-Holder_Arxiv_2009, Bussmann-Holder_CondMat_2019, Gupta_FrontPhys_2019, Khasanov_AuBe_PRR_2020, Khasanov_AuBe_PRB_2020}, within the two-gap approach, the coupled $s$-wave gap equations are described as:
\begin{equation}
\Delta_i (T)= \sum_{j=1}^2 \int_{0}^{\omega_{D_j}}\frac{n_jV_{ij}\Delta_j(T)}{\sqrt{\epsilon^2+\Delta_j^2(T)}}\tanh \frac{\sqrt{\epsilon^2+\Delta_j^2(T)}}{2k_{\rm B}T}d\epsilon
 \label{eq:Self-consistent_model}
\end{equation}
Here $n_i$ is the partial density of states for $i-$th band at the Fermi level ($n_1+n_2=1$); $V_{ii}$ and $V_{ij\neq i}$ are the intraband and interband interaction potentials, respectively. For simplicity, one normally assumes that the Debye frequency $\omega_{\rm D}$ is the same for both bands ($\omega_{\rm D1}=\omega_{\rm D2}=\omega_{\rm D}$).\cite{Kogan_PRB_2009, Bussmann-Holder_Arxiv_2009, Khasanov_PRL_2010, Gupta_FrontPhys_2019, Khasanov_AuBe_PRR_2020, Khasanov_AuBe_PRB_2020} The system of self-consistent coupled gap equations (Eq.~\ref{eq:Self-consistent_model}) might be solved for the certain values of the intraband ($V_{11}$, $V_{22}$) and interband ($V_{12}$, $V_{21}$) coupling potentials and the Debye frequency ($\omega_{\rm D}$), which results in temperature dependencies of the big $\Delta_1(T)$ and the small $\Delta_2(T)$ superconducting energy gaps.

The thermodynamic critical field $B_{\rm c}$ is further obtained as the difference between the normal state (NS) and the superconducting (SC) free energies ($F_{\rm NS}$ and $F_{\rm SC}$) \cite{Khasanov_AuBe_PRR_2020, Khasanov_AuBe_PRB_2020}:
\begin{eqnarray}
\frac{B_{\rm c}^2(T)}{\gamma_{\rm e}T_{\rm c}^2} &=& 8 \pi \frac{F_{\rm NS}-F_{\rm SC1}-F_{\rm SC2}}{\gamma_{\rm e}T_{\rm c}^2} \nonumber \\
       &=& 8 \pi \left[ \frac{F_{\rm NS}}{\gamma_{\rm e}T_{\rm c}^2} - \omega_{F_{\rm s}}\frac{F_{\rm SC1}}{\gamma_{\rm e1}T_{\rm c}^2} -(1-\omega_{F_{\rm s}})\frac{F_{\rm SC2}}{\gamma_{\rm e2}T_{\rm c}^2} \right] . \nonumber \\
       &&
  \label{eq:Critical-field_two-gap}
\end{eqnarray}
$F_{\rm NS}$ and $F_{\rm SC}$ are determined as:\cite{Johnston_SST_2013,Padamsee_JLTP_1973}
\begin{equation}
\frac{F_{\rm NS}}{\gamma_{\rm e}T_{\rm c}^2} = -\frac{ T^2}{2 T_{\rm c}^2}.
 \label{eq:Fn}
\end{equation}
and
\begin{eqnarray}
\frac{F_{\rm SC}[T,\Delta(T)]}{\gamma_{\rm e}T_{\rm c}^2}  & = &-\frac{3 }{ \pi^2 k_{\rm B}^2 T_{\rm c}^2}
\bigg[  \frac{\Delta(T)^2}{4}  \bigg. \nonumber \\
&& + \left.  \int_0^\infty f\; \frac{2\epsilon^2 + \Delta(T)^2}{\sqrt{\epsilon^2 + \Delta(T)^2}} d\epsilon \right],
\label{eq:Fs}
\end{eqnarray}
respectively, where $f=[1+\exp(\sqrt{\epsilon^2+\Delta(T)^2}/k_{\rm B}T)]^{-1}$ is the Fermi function and $\gamma_{\rm e}$ is the electronic specific heat component.

The contribution of the bigger gap to the Free energy ($\omega$) is determined by the normal state electronic specific heat $\gamma_{\rm e}$ ($\gamma_{\rm e}=\gamma_{\rm e1}+\gamma_{\rm e2}$) and is equal to \cite{Khasanov_AuBe_PRB_2020}:
\begin{equation}
\omega = \omega_{F_{\rm s}}=\frac{n_1}{n_1+n_2}= \frac{\gamma_{e1}}{\gamma_{e1}+\gamma_{e2}}.
 \label{eq:omega_Ce_Bc}
\end{equation}

\subsection{Approximate equations for the electron-phonon interaction}

To obtain simplified expressions for $\lambda$ and $\omega_{log}$ we make the following assumptions:

\begin{enumerate}
	\item Au and Be vibrational modes are completely disentangled \textcolor{black}{(justified by well-separated Au and Be phonon DOS in Fig. \ref{fig:phdos}).}
	\item \textcolor{black}{Electron-phonon coupling with Au phonons} is negligible.
	\item \textcolor{black}{Electron-phonon coupling with Be phonons is constant (justified by the narrow Be phonon DOS in Fig.~\ref{fig:phdos}).}
	\item The electron-phonon interaction is isotropic.
\end{enumerate}

\noindent Following the convention in Ref. \cite{ReviewH}, $\lambda$ and $\omega_{\log}$ are given by:

\begin{equation}
	\lambda = 2 \int d \omega \frac
	{\alpha^2 F(\omega)} {\omega}
	\label{eq:lambda}
\end{equation}

\begin{equation}
	\omega_{log} = \exp
	\left[
		\frac 2 {\lambda}
		\int d\omega \log(\omega) \frac {\alpha^2 F (\omega)} {\omega}
	\right],
	\label{eq:omegalog}
\end{equation}

\noindent where $\alpha^2 F$ is given by Eq. 15, 53 of Ref. \cite{ReviewH}. Here, we simplify the notation by writing:

\begin{equation}
	\alpha^2F(\omega) = \frac {\hbar} {N(E_f)}
	\sum_{\mathbf q\nu}
	\frac {f_{\nu}(\mathbf q)} {2 \omega_{\mathbf q \nu}}
	\delta(\omega - \omega_{\mathbf q \nu}),
	\label{eq:alfa2f}
\end{equation}

\noindent where we introduce a function $f_{\nu}(\mathbf q)$ containing all the information about the electronic structure and about the bare electron-phonon coupling constant. Its explicit form can be written down combining Eq. 15 and 53 of the same reference. The index $\nu$ specifies the phonon branch (or 'band' index), while $\mathbf q$ is the wave-vector. The pre-factor $N(E_f)$ is the electronic DOS at the Fermi level.

Inserting this expression into Eq. \ref{eq:lambda} and \ref{eq:omegalog}, and performing the integral over the variable $\omega$ to exploit the Dirac-delta appearing in \ref{eq:alfa2f}, we get:
\begin{equation}
	\lambda = \frac {\hbar} {N(E_f)}
	\sum_{\mathbf q\nu}
	\frac {f_{\nu}(\mathbf q)} {\omega_{\mathbf q \nu}^2}
	\quad \text{and}
\end{equation}
\begin{equation}
	\omega_{log} = \exp
	\left[
		\frac {\hbar} {\lambda N(E_f)}
		\sum_{\mathbf q \nu}
		\log(\omega_{\mathbf q\nu})
		\frac {f(\omega_{\mathbf q\nu})} {\omega_{\mathbf q \nu}^{2}}
	\right].
\end{equation}

\noindent Considering the electron-phonon interaction to be isotropic, we expect $f_{\nu}(\mathbf q)$ to depend mainly on the absolute value of $q$ rather than its specific direction. In other words, we expect $f_{\nu}(\mathbf q)$ to explicitly depend on the phonon energies, rather than $q$-vector. Introducing the phonon density of states $\rho(\omega)$ we can, therefore, replace the summation over $\mathbf q$ and $\nu$ with an integral over $\omega$:
\begin{equation}
	\lambda = \frac {\hbar} {N(E_f)}
	\int d\omega \rho (\omega )
	\frac {f(\omega)} {\omega^2}
\end{equation}
\begin{equation}
	\omega_{log} = \exp
	\left[
		\frac {\hbar} {\lambda N(E_f)}
		\int d\omega \rho (\omega )
		\frac {f(\omega)} {\omega^2}
		\log(\omega)
	\right].
\end{equation}

\noindent Finally, if we assume Au contribution to be negligible, we can start the integral from some frequency $\omega_1$ above 200 cm$^{-1}$ and regard $f$ to depend only the optic Be-induced modes. Then, in view of the small energy dispersion (Fig. \ref{fig:phdos}) it is safe to assume $f$ to be constant within this energy range and bring it outside of the integral. Defining the proportionality constant:

\begin{equation}
	\gamma \equiv \frac {\hbar f} {N(E_f)},
\end{equation}
we obtain Eq. \ref{eq:lambda} in the main text. For $\omega_{log}$ the dependence on $\gamma$ drops out and we have:
\begin{equation}
	\omega_{log} = \exp
	\left[
		\frac {
		\int d\omega \rho (\omega )
		\frac {f(\omega)} {\omega^2}
		\log(\omega)}
		{
		\int d\omega \rho (\omega )
		\frac {f(\omega)} {\omega^2}
		}
	\right].
\end{equation}

\noindent Since $N(E_f)$ is constant in the pressure range here considered (see Fig. \ref{fig:edos} in the main text), we are justified to estimate $\gamma$ from the ambient pressure measurement of $T_{\rm c}$ (see main text).


\begin{thebibliography}{999}

\bibitem{Sigrist_AIP-Cof-Proc_2009} Manfred Sigrist, {\it Introduction to unconventional superconductivity in non‐centrosymmetric metals}, AIP Conf. Proc. {\bf 1162}, 55 (2009).\\
    \url{https://doi.org/10.1063/1.3225489}

\bibitem{Yip_AnRevCondMat_2014} Sungkit Yip, {\it Noncentrosymmetric Superconductors}, Annu. Rev. Condens. Matter Phys. {\bf 5}, 15 (2014).\\
    \url{https://doi.org/10.1146/annurev-conmatphys-031113-133912}

\bibitem{Smidman_RPP_2017} M. Smidman, M. B. Salamon, H. Q. Yuan, D. F. Agterberg, {\it Superconductivity and spin-orbit coupling in non-centrosymmetric materials: a review}, Rep. Prog. Phys. {\bf 80}, 036501 (2017).\\
    \url{https://doi.org/10.1088/1361-6633/80/3/036501}

\bibitem{Takimoto_JPSJ_2009} T. Takimoto and P. Thalmeier, {\it Triplet Cooper Pair Formation by Anomalous Spin Fluctuations in Non-centrosymmetric Superconductors}, J. Phys. Soc. Jpn. {\bf 78}, 103703 (2009).\\
    \url{https://doi.org/10.1143/JPSJ.78.103703}

\bibitem{Shyta} V. Shyta, Jeroen van den Brink, Flavio S Nogueira, {\it Chiral Meissner state in time-reversal invariant Weyl superconductors}, Physical Review Research {\bf 6}, 013240 (2024).\\
    \url{https://doi.org/10.1103/PhysRevResearch.6.013240}

\bibitem{Matthis_JPCS_1959} B. Matthias, {\it Superconductivity of AuBe}, J. Phys. Chem. Solids {\bf 10}, 342 (1959).\\
    \url{https://doi.org/10.1016/0022-3697(59)90015-0}

\bibitem{Rebar_thesis_2015} D. J. Rebar, LSU Doctoral Dissertations 3989 (2015).\\ \url{https:digitalcommons.lsu.edu/gradschool_dissertations/3989/}

\bibitem{Amon_PRB_2018} A. Amon, E. Svanidze, R. Cardoso-Gil, M. N. Wilson, H. Rosner, M. Bobnar, W. Schnelle, J. W. Lynn, R. Gumeniuk, C. Hennig, G. M. Luke, H. Borrmann, A. Leithe-Jasper, and Yu. Grin, {\it Noncentrosymmetric superconductor BeAu}, Phys. Rev. B {\bf 97}, 014501 (2018). \\
    \url{https://doi.org/10.1103/PhysRevB.97.014501}

\bibitem{Singh_PRB_2019} D. Singh, A. D. Hillier, and R. P. Singh, {\it Type-I superconductivity in the noncentrosymmetric superconductor BeAu},  Phys. Rev. B {\bf 99}, 134509 (2019).\\
    \url{https://doi.org/10.1103/PhysRevB.99.134509}

\bibitem{Beare_PRB_2019} J. Beare, M. Nugent, M.N. Wilson, Y. Cai, T.J.S. Munsie, A. Amon, A. Leithe-Jasper, Z. Gong, S.L. Guo, Z. Guguchia, Y. Grin, Y.J. Uemura, E. Svanidze, and G.M. Luke, {\it $\mu$SR and magnetometry study of the type-I superconductor BeAu}, Phys. Rev. B {\bf 99}, 134510 (2019).\\
    \url{https://doi.org/10.1103/PhysRevB.99.134510}

\bibitem{Rebar_PRB_2019} Drew J. Rebar, Serena M. Birnbaum, John Singleton, Mojammel Khan, J. C. Ball, P. W. Adams, Julia Y. Chan, D. P. Young, Dana A. Browne, and J. F. DiTusa, {\it Fermi surface, possible unconventional fermions, and unusually robust resistive critical fields in the chiral-structured superconductor AuBe}, Phys. Rev. B {\bf 99}, 094517 (2019). \\
    \url{https://doi.org/10.1103/PhysRevB.99.094517}

\bibitem{Khasanov_AuBe_PRR_2020} Rustem Khasanov, Ritu Gupta, Debarchan Das, Alfred Amon, Andreas Leithe-Jasper, and Eteri Svanidze, {\it     Multiple-gap response of type-I noncentrosymmetric BeAu superconductor}, Phys. Rev. Research {\bf 2}, 023142 (2020). \\
    \url{https://doi.org/10.1103/PhysRevResearch.2.023142}

\bibitem{Khasanov_AuBe_PRB_2020} Rustem Khasanov, Ritu Gupta, Debarchan Das, Andreas Leithe-Jasper, and Eteri Svanidze, {\it Single-gap versus two-gap scenario: Specific heat and thermodynamic critical field of the noncentrosymmetric superconductor BeAu}, Phys. Rev. B {\bf 102}, 014514 (2020). \\
    \url{https://doi.org/10.1103/PhysRevB.102.014514}

\bibitem{Datta_PRB_2022} Soumya Datta, Aastha Vasdev, Partha Sarathi Rana, Kapil Motla, Anshu Kataria, Ravi Prakash Singh, Tanmoy Das, and Goutam Sheet, {\it Spectroscopic evidence of multigap superconductivity in noncentrosymmetric AuBe}, Phys. Rev. B {\bf 105}, 104505 (2022). \\
    \url{https://doi.org/10.1103/PhysRevB.105.104505}

\bibitem{Peets} D. Peets, E. Cheng, T. Ying, M. Kriener, X. Shen, S. Li, and D. Feng, {\it Type-I superconductivity in Al$_6$Re}, Phys. Rev. B {\bf 99}, 14 (2019).\\
    \url{https://doi.org/10.1103/PhysRevB.99.144519}

\bibitem {lv2020a} B. Lv, M. Li, J. Chen, Y. Yang, S. Wu, L. Qiao, F. Guan, H. Xing, Q. Tao, G. Cao, and Z. Xu, {\it Type-i superconductivity in noncentrosymmetric NbGe$_2$}, Phys. Rev. B {\bf 102}, 64507 (2020).\\
    \url{https://doi.org/10.1103/PhysRevB.102.064507}

\bibitem{svanidze2012a} E. Svanidze and E. Morosan], {\it Type-I superconductivity in ScGa$_3$ and LuGa$_3$ single crystals}, Phys. Rev. B {\bf 85}, 174514 (2012).\\
    \url{https://doi.org/10.1103/PhysRevB.85.174514}

\bibitem{zhao2012a} L. Zhao, S. Lausberg, H. Kim, M. Tanatar, M. Brando, R. Prozorov, and E. Morosan, {\it Type-I superconductivity in YbSb$_2$ single crystals}, Phys. Rev. B {\bf 85}, 214526 (2012).\\
    \url{https://doi.org/10.1103/PhysRevB.85.214526}

\bibitem{takeda2015a} M. Takeda, A. Teruya, A. Nakamura, H. Harima, M. Hedo, T. Nakama, and Y. Onuki, {\it De haas-van alphen effect in Rh$_2$Ga$_9$ and Ir$_2$Ga$_9$ without inversion symmetry in the crystal structure and related compounds T$_2$Al$_9$ (T:Co, Rh, Ir) with inversion symmetry}, J. Phys. Soc. Japan {\bf 84}, 2 (2015).\\
    \url{https://doi.org/10.7566/JPSJ.84.024701}

\bibitem{biswas2020a} P. Biswas, F. Rybakov, R. Singh, S. Mukherjee, N. Parzyk, G. Balakrishnan, M. Lees, C. Dewhurst, E. Babaev, A. Hillier, and D. Paul, {\it Coexistence of type-I and type-II superconductivity signatures in ZrB$_{12}$ probed by muon spin rotation measurements}, Phys. Rev. B {\bf 102}, 144523 (2020).\\
    \url{https://doi.org/10.1103/PhysRevB.102.144523}

\bibitem{anand2011a} V. Anand, A. Hillier, D. Adroja, A. Strydom, H. Michor, K. McEwen, and B. Rainford, {\it Specific heat and $\mu$SR study on the noncentrosymmetric superconductor LaRhSi$_3$}, Phys. Rev. B {\bf 83}, 64522 (2011).\\
    \url{https://doi.org/10.1103/PhysRevB.83.064522}
	
\bibitem{smidman2014a} M. Smidman, A. Hillier, D. Adroja, M. Lees, V. Anand, R. Singh, R. Smith, D. Paul, and G. Balakrishnan, {\it Investigations of the superconducting states of noncentrosymmetric LaPdSi$_3$ and LaPtSi$_3$}, Phys. Rev. B {\bf 89}, 94509 (2014).\\
    \url{https://doi.org/10.1103/PhysRevB.89.094509}
	
\bibitem{yonezawa2005a} S. Yonezawa and Y. Maeno, {Type-I superconductivity of the layered silver oxide Ag$_5$Pb$_2$O$_6$}, Phys. Rev. B {\bf 72}, 18 (2005).\\
    \url{https://doi.org/10.1103/PhysRevB.72.180504}
	
\bibitem{hull1981a} G. Hull, J. Wernick, T. Geballe, J. Waszczak, and J. Bernardini, {\it Superconductivity in the ternary intermetallics YbPd$_2$Ge$_2$, LaPd$_2$Ge$_2$, and LaPt$_2$Ge$_2$}, Phys. Rev. B {\bf 24}, 6715 (1981).\\
    \url{https://doi.org/10.1103/PhysRevB.24.6715}
	
\bibitem{palstra1986a} T. Palstra, G. Lu, A. Menovsky, G. Nieuwenhuys, P. Kes, and J. Mydosh, {\it Superconductivity in the ternary rare-earth (Y, La, and Lu) compounds RPd$_2$Si$_2$ and RRh$_2$Si$_2$}, Phys. Rev. B {\bf 34}, 4566 (1986).\\
    \url{https://doi.org/10.1103/PhysRevB.34.4566}

\bibitem{Leng_PRB_2017} H. Leng, C. Paulsen, Y. K. Huang, and A. de Visser, {\it Type-I superconductivity in the Dirac semimetal PdTe$_2$}, Phys. Rev. B {\bf 96}, 220506(R) (2017). \\
    \url{https://doi.org/10.1103/PhysRevB.96.220506}
	
\bibitem{g2020a} K. G\'{o}rnicka, G. Kuderowicz, E. Carnicom, K. Kutorasi\'{n}ski, B. Wiendlocha, R. Cava, and T. Klimczuk, {\it Soft-mode enhanced type-I superconductivity in LiPd$_2$Ge}, Phys. Rev. B {\bf 102}, 24507 (2020).\\
    \url{https://doi.org/10.1103/PhysRevB.102.024507}
	
\bibitem{nakamura2013a} A. Nakamura, H. Harima, M. Hedo, T. Nakama, and Y. Onuki, {\it Chiral crystal structure and split Fermi surface properties in TaSi$_2$}, J. Phys. Soc. Japan {\bf 82}, 113705 (2013).\\
    \url{https://doi.org/10.7566/JPSJ.82.113705}
	
\bibitem{herrmannsdoerfer1996a} T. Herrmannsd\"{o}rfer, S. Rehmann, and F. Pobell, {\it Interplay between nuclear ferromagnetism and superconductivity in AuIn$_2$}, Czechoslov. J. Phys {\bf 46}, 2189 (1996).\\
    \url{https://doi.org/10.1007/BF02571086}
	
\bibitem{kobayashi1981a} M. Kobayashi and I. Tsujikawa, {\it Potassium concentration dependence of the superconductivity in the potassium graphite intercalation compounds}, J. Phys. Soc. Japan {\bf 50}, 3245 (1981).\\
    \url{https://doi.org/10.1143/JPSJ.50.3245}
	
\bibitem{sun2016a} S. Sun, K. Liu, and H. Lei, {\it Type-I superconductivity in KBi$_2$ single crystals}, J. Phys. Condens. Matter {\bf 28}, 085701 (2016).\\
    \url{https://doi.org/10.1088/0953-8984/28/8/085701}
	
\bibitem{bekaert2016a} J. Bekaert, S. Vercauteren, A. Aperis, L. Komendov\'{a}, R. Prozorov, B. Partoens, and M. V. Milo\v{s}evic, {\it Anisotropic type-I superconductivity and anomalous superfluid density in OsB$_2$}, Phys. Rev. B {\bf 94}, 144506 (2016).\\
    \url{https://doi.org/10.1103/PhysRevB.94.144506}
	
\bibitem{ren2007a} Z. Ren, J. Kato, T. Muranaka, J. Akimitsu, M. Kriener, and Y. Maeno, {\it Superconductivity in boron-doped SiC}, J. Phys. Soc. Japan {\bf 76}, 10 (2007).\\
    \url{https://doi.org/10.1143/JPSJ.76.103710}
	
\bibitem{wang2014a} Y. Wang, H. Sato, Y. Toda, S. Ueda, H. Hiramatsu, and H. Hosono, {\it SnAs with the NaCl-type structure: Type-I superconductivity and single valence state of Sn}, Chem. Mater. {\bf 26}, 7209 (2014).\\
    \url{https://doi.org/10.1021/cm503992d}

\bibitem{Arushi_PRB_2021} Arushi, K. Motla, A. Kataria, S. Sharma, J. Beare, M. Pula, M. Nugent, G. M. Luke, and R. P. Singh, {\it Type-I superconductivity in single-crystal Pb$_2$Pd}, Phys. Rev. B {\bf 103}, 184506 (2021).\\
    \url{https://doi.org/10.1103/PhysRevB.103.184506}

\bibitem{Lorenz_2005} B. Lorenz and C. Chu,  {\it High Pressure Effects on Superconductivity}, In: Narlikar, A.V. (eds) Frontiers in Superconducting Materials. Springer, Berlin, Heidelberg. (2005). \\
    \url{https://doi.org/10.1007/3-540-27294-1_12}

\bibitem{Schilling_book_2007} J.S. Schilling, {\it High-Pressure Effects}, In: Schrieffer, J.R., Brooks, J.S. (eds) Handbook of High-Temperature Superconductivity. Springer, New York, NY. (2007). \\
    \url{https://doi.org/10.1007/978-0-387-68734-6_11}

\bibitem{Schilling_JPCS_2008} J.S. Schilling and J. J. Hamlin, {\it Recent studies in superconductivity at extreme pressures}, J. Phys.: Conf. Ser. {\bf 121}, 052006 (2008).\\
    \url{https://doi.org/10.1088/1742-6596/121/5/052006}

\bibitem{Leithe-Jasper_MPI-report_2003} A. Leithe-Jasper, H. Borrmann, and W. H\"{o}nle, in Max Plank Institute for Chemical Physics of Solids, Scientific Report (Dresden, 2003–2005), p.~25.

\bibitem{Topas} Topas 7, {\it Bruker AXS}.\\
    \url{https://www.brukersupport.com/ProductDetail/1145} 

\bibitem{Willmott_JSyncRad_2013} P. Willmott, D. Meister, S. Leake, M. Lange, A. Bergamaschi, M. B\"{o}ge, M. Calvi, C. Cancellieri, N. Casati, A. Cervellino, Q. Chen, C. David, U. Flechsig, F. Gozzo, B. Henrich, S. J\"{a}ggi-Spielmann, B. Jakob, I. Kalichava, P. Karvinen, J. Krempasky,  A L\"{u}deke, R L\"{u}scher, S Maag, C Quitmann, M L Reinle-Schmitt, T Schmidt, B Schmitt, A Streun, I Vartiainen, M Vitins, X Wang, and R Wullschleger, {\it The Materials Science beamline upgrade at the Swiss Light Source}, J. Synchrotron Radiat. {\bf 20}, 667 (2013).\\
    \url{https://doi.org/10.1107/S0909049513018475}

\bibitem{Angel_1993} R. J. Angel, {\it The high-pressure, high-temperature equation of state of calcium fluoride, CaF$_2$}, Journal of Physics: Condensed Matter {\bf 5}, L141 (1993).\\
    \url{https://doi.org/10.1088/0953-8984/5/11/001}

\bibitem{Daphne} D. Sta{\v{s}}ko, J. Prchal, M. Klicpera, S. Aoki, and K. Murata, {\it Pressure media for high pressure experiments, Daphne Oil 7000 series}, High Pressure Research {\bf 40}, 525 (2020).\\
    \url{https://doi.org/10.1088/0953-8984/5/11/001}

\bibitem{Khasanov_JAP_2022} Rustem Khasanov, {\it Perspective on muon-spin rotation/relaxation under hydrostatic pressure}, J. Appl. Phys. {\bf 132}, 190903 (2022). \\
    \url{https://doi.org/10.1063/5.0119840}

\bibitem{Khasanov_HPR_2016} R. Khasanov, Z. Guguchia, A. Maisuradze, D. Andreica, M. Elender, A. Raselli, Z. Shermadini, T. Goko, F. Knecht, E. Morenzoni, and A. Amato, {\it High pressure research using muons at the Paul Scherrer Institute}, High Pressure Res. {\bf 36}, 140 (2016).\\
    \url{https://doi.org/10.1080/08957959.2016.1173690}

\bibitem{MUSRFIT} A. Suter and B. M. Wojek, {\it Musrfit: A Free Platform-Independent Framework for $\mu$SR Data Analysis}, Phys. Procedia {\bf 30}, 69 (2012).\\
    \url{https://doi.org/10.1016/j.phpro.2012.04.042}

\bibitem{Naumov_PRA_2022} Pavel Naumov, Ritu Gupta, Marek Bartkowiak, Ekaterina Pomjakushina, Nicola P.M. Casati, Matthias Elender, and Rustem Khasanov, {\it Optical Setup for a Piston-Cylinder Pressure Cell: A Two-Volume Approach}, Phys. Rev. Applied {\bf 17}, 024065 (2022).\\
    \url{https://doi.org/10.1103/PhysRevApplied.17.024065}

\bibitem{FPLO} K. Koepernik and H. Eschrig, {\it Full-potential nonorthogonal local-orbital minimum-basis band-structure scheme}, Phys. Rev. B {\bf 59}, 1743 (1999).\\
    \url{https://doi.org/10.1103/PhysRevB.59.1743}
	
\bibitem{PBE} J. P. Perdew, K. Burke, and M. Ernzerhof, {\it Generalized gradient approximation made simple}, Phys. Rev. Lett. {\bf 77}, 3865 (1996).\\
    \url{https://doi.org/10.1103/PhysRevLett.78.1396}
	
\bibitem{QE} P. Giannozzi, S. Baroni, N. Bonini, M. Calandra, R. Car, C. Cavazzoni, D. Ceresoli, G.~L. Chiarotti, M. Cococcioni, I. Dabo, A.~D. Corso, S. de~Gironcoli, S. Fabris, G. Fratesi, R. Gebauer, U. Gerstmann, C. Gougoussis, A. Kokalj, M. Lazzeri, L. Martin-Samos, N. Marzari, F. Mauri, R. Mazzarello, S. Paolini, A. Pasquarello, L. Paulatto, C. Sbraccia, S. Scandolo, G. Sclauzero, A.~P. Seitsonen, A. Smogunov, P. Umari, and R.~M. Wentzcovitch, {\it Quantum espresso: a modular and open-source software project for quantum simulations of materials}, J. Phys. Condens. Matter {\bf 21}, 395502 (2009).\\
    \url{https://doi.org/10.1088/0953-8984/21/39/395502}
	
\bibitem{QE2} P. Giannozzi, O. Andreussi, T. Brumme, O. Bunau, M.~B. Nardelli, M. Calandra, R. Car, C. Cavazzoni, D. Ceresoli, M. Cococcioni, N. Colonna, I. Carnimeo, A.~D. Corso, S. de~Gironcoli, P. Delugas, R.~A. DiStasio, A. Ferretti, A. Floris, G. Fratesi, G. Fugallo, R. Gebauer, U. Gerstmann, F. Giustino, T. Gorni, J. Jia, M. Kawamura, H.-Y. Ko, A. Kokalj, E. Küçükbenli, M. Lazzeri, M. Marsili, N. Marzari, F. Mauri, N.~L. Nguyen, H.-V. Nguyen, A.~O. de-la Roza, L. Paulatto, S. Poncé, D. Rocca, R. Sabatini, B. Santra, M. Schlipf, A.~P. Seitsonen, A. Smogunov, I. Timrov, T. Thonhauser, P. Umari, N. Vast, X. Wu, and S. Baroni, {\it Advanced capabilities for materials modelling with quantum espresso}, J. Phys. Condens. Matter {\bf 29}, 465901 (2017).\\
    \url{https://doi.org/10.1088/1361-648X/aa8f79}

\bibitem{Strassle_PRB_2014} Th. Str\"{a}ssle, S. Klotz, K. Kunc, V. Pomjakushin, and J. S. White, {\it Equation of state of lead from high-pressure neutron diffraction up to 8.9 GPa and its implication for the NaCl pressure scale}, Phys. Rev. B {\bf 90}, 014101 (2014).\\
    \url{https://doi.org/10.1103/PhysRevB.90.014101}

\bibitem{Klotz_PRB_2017} S. Klotz, K. Komatsu, H. Kagi, K. Kunc, A. Sano-Furukawa, S. Machida, and T. Hattori, {\it Bulk moduli and equations of state of ice VII and ice VIII}, Phys. Rev. B {\bf 95}, 174111 (2017).\\ \url{https://doi.org/10.1103/PhysRevB.95.174111}

\bibitem{Murnaghan_AmJMath_1937} F. D. Murnaghan, {\it Finite Deformations of an Elastic Solid}, Am. J. Math. {\bf 59}, 235 (1937).\\
    \url{https://doi.org/10.2307/2371405}

\bibitem{Eiling_JPF_1981} A. Eiling and J. S. Schilling, {\it Pressure and temperature dependence of electrical resistivity of Pb and Sn from 1-300 K and 0-10 GPa-use as continuous resistive pressure monitor accurate over wide temperature range; superconductivity under pressure in Pb, Sn and In}, J. Phys. F: Met. Phys. {\bf 11}, 623 (1981).\\
    \url{https://doi.org/10.1088/0305-4608/11/3/010}

\bibitem{Bardeen_PR_1957} J. Bardeen, L.~N. Cooper, and J.~R. Schrieffer, {\it Theory of superconductivity}, Phys. Rev. {\bf 108}, 1175 (1957).\\
    \url {https://doi.org/10.1103/PhysRev.108.1175}


	
\bibitem{EliashbergReview} F. Marsiglio, {\it Eliashberg theory: A short review}, Annals of Physics {\bf 417}, 168102 (2020).\\
    \url{https://doi.org/https://doi.org/10.1016/j.aop.2020.168102}
	
\bibitem{AllenDynes} P.~B. Allen and R.~C. Dynes, {\it Transition temperature of strong-coupled superconductors reanalyzed}, Phys. Rev. B {\bf 12}, 905 (1975).\\
    \url{https://doi.org/10.1103/PhysRevB.12.905}
	
\bibitem{ReviewH} J.~A. Flores-Livas, L. Boeri, A. Sanna, G. Profeta, R. Arita, and M. Eremets, {\it A perspective on conventional high-temperature superconductors at high pressure: Methods and materials}, Physics Reports {\bf 856}, 1 (2020).\\
    \url{https://doi.org/https://doi.org/10.1016/j.physrep.2020.02.003}

\bibitem{Huebener_book_2001} R. P. Huebener, {\it Magnetic Flux Structures in Superconductors}, Springer Series in Solid-State Sciences, Springer Berlin, Heidelberg. (2001).\\
    \url{https://doi.org/10.1007/978-3-662-02305-1}

\bibitem{Karl_PRB_2019} Richard Karl, Florence Burri, Alex Amato, Mauro Donegà, Severian Gvasaliya, Hubertus Luetkens, Elvezio Morenzoni, and Rustem Khasanov, {\it Muon spin rotation study of type-I superconductivity: Elemental $\beta-$Sn}, Phys. Rev. B {\bf 99}, 184515 (2019). \\
    \url{https://doi.org/10.1103/PhysRevB.99.184515}

\bibitem{Khasanov_BiII_PRB_2019} Rustem Khasanov, Milo\u{s} M. Radonji\'{c}, Hubertus Luetkens, Elvezio Morenzoni, Gediminas Simutis, Stephan Sch\"{o}necker, Wilhelm H. Appelt, Andreas \"{O}stlin, Liviu Chioncel, and Alex Amato, {\it Superconducting nature of the Bi-II phase of elemental bismuth}, Phys. Rev. B {\bf 99}, 174506 (2019). \\
    \url{https://doi.org/10.1103/PhysRevB.99.174506}

\bibitem{Khasanov_PRL_2010} R. Khasanov, M. Bendele, A. Amato, K. Conder, H. Keller, H.-H. Klauss, H. Luetkens, and E. Pomjakushina, {\it Evolution of Two-Gap Behavior of the Superconductor FeSe$_{1-x}$}, Phys. Rev. Lett. {\bf 104}, 087004 (2010).\\
     \url{https://doi.org/10.1103/PhysRevLett.104.087004}

\bibitem{Khasanov_PRB_2011} R. Khasanov, S. Sanna, G. Prando, Z. Shermadini, M. Bendele, A. Amato, P. Carretta, R. De Renzi, J. Karpinski, S. Katrych, H. Luetkens, and N. D. Zhigadlo, {\it Tuning of competing magnetic and superconducting phase volumes in LaFeAsO$_{0.945}$F$_0.055$ by hydrostatic pressure}, Phys. Rev. B {\bf 84}, 100501(R) (2011).\\
    \url{https://doi.org/10.1103/PhysRevB.84.100501}

\bibitem{Yaouanc_book_2011} A. Yaouanc, and P. Dalmas de R\'{e}otier, {\it Muon Spin Rotation, Relaxation and Resonance: Applications to Condensed Matter} (Oxford University Press, Oxford, 2011).

\bibitem{Padamsee_JLTP_1973} H. Padamsee, J. E. Neighbor, and C. A. Shiffman, {\it Quasiparticle phenomenology for thermodynamics of strong-coupling superconductors},  J. Low Temp. Phys. {\bf 12}, 387 (1973).\\
    \url{https://doi.org/10.1007/BF00654872}

\bibitem{Johnston_SST_2013} D.C. Johnston, {\it Elaboration of the $\alpha-$model derived from the BCS theory of superconductivity}, Supercond. Sci. Technol. {\bf 26}, 115011 (2013).\\
    \url{https://doi.org/10.1088/0953-2048/26/11/115011}

\bibitem{Mathcad} \url{https://www.mathcad.com/de}

\bibitem{Gonnelli_PRL_2002} R. S. Gonnelli, D. Daghero, G. A. Ummarino, V. A. Stepanov, J. Jun, S. M. Kazakov, and J. Karpinski, {\it Direct Evidence for Two-Band Superconductivity in MgBB$_2$ Single Crystals from Directional Point-Contact Spectroscopy in Magnetic Fields}, Phys. Rev. Lett. {\bf 89}, 247004 (2002). \\
    \url{https://doi.org/10.1103/PhysRevLett.89.247004}

\bibitem{Iavarone_PRL_2002} M. Iavarone, G. Karapetrov, A. E. Koshelev, W. K. Kwok, G. W. Crabtree, D. G. Hinks, W. N. Kang, Eun-Mi Choi, Hyun Jung Kim, Hyeong-Jin Kim, and S. I. Lee, {\it Two Band Superconductivity in MgB$_2$}, Phys. Rev. Lett. {\bf 89}, 187002 (2002). \\
    \url{https://doi.org/10.1103/PhysRevLett.89.187002}

\bibitem{Chen_APL_2008} Ke Chen, Y. Cui, Qi Li, C. G. Zhuang, Zi-Kui Liu, and X. X. Xi, {\it Study of MgB$_2$/I/Pb tunnel junctions on MgO (211) substrates}, Appl. Phys. Lett. {\bf 93}, 012502 (2008). \\
    \url{https://doi.org/10.1063/1.2956414}

\bibitem{Kim_Symmetry_2019} H. Kim, K. Cho, M.A. Tanatar, V. Taufour, S.K. Kim, S.L. Bud’ko, P.C. Canfield, V.G. Kogan, and R. Prozorov, {\it Self-Consistent Two-Gap Description of MgB$_2$ Superconductor}, Symmetry {\bf 11}, 1012 (2019).\\
    \url{ https://doi.org/10.3390/sym11081012}

\bibitem{Tinkham_book_1975} M. Tinkham, {\it Introduction to Superconductivity} (Krieger Publishing company,
    Malabar, Florida, 1975).

\bibitem{Kogan_PRB_2009} V. G. Kogan, C. Martin, and R. Prozorov, {\it Superfluid density and specific heat within a self-consistent scheme for a two-band superconductor}, Phys. Rev. B {\bf 80}, 014507 (2009).\\
    \url{https://doi.org/10.1103/PhysRevB.80.014507}

\bibitem{Gupta_FrontPhys_2019} R. Gupta, A. Maisuradze, N. D. Zhigadlo, H. Luetkens, A. Amato, and R. Khasanov, {\it Self-Consistent Two-Gap Approach in Studying Multi-Band Superconductivity of NdFeAsO$_{0.65}$F$_{0.35}$}, Front. Phys. {\bf 8}, 2 (2020).\\
    \url{https://doi.org/10.3389/fphy.2020.00002}

\bibitem{Eliashberg_1960a} G. M. Eliashberg, {\it Intercations between Electronic and Lattice Vibrations in a Superconductor}, Sov. Phys. JETP {\bf 11}, 696 (1960).

\bibitem{Eliashberg_1960b} G. M. Eliashberg, Zh. Eksp. Teor. Fiz. {\bf 38}, 966 (1960).

\bibitem{Carbotte_RMP_1990} J.P. Carbotte, {\it Properties of boson-exchange superconductors}, Rev. Mod. Phys. {\bf 62}, 1027 (1990).\\
    \url{https://doi.org/10.1103/RevModPhys.62.1027}

\bibitem{Marsiglio_book_2008} F. Marsiglio and J. P. Carbotte, 'Electron-Phonon Superconductivity`, Review Chapter in {\it Superconductivity, Conventional and Unconventional Superconductors}, edited by K. H. Bennemann and J. B. Ketterson (Springer-Verlag, Berlin, 2008).

\bibitem{Scalapino_book_1969} D.J. Scalapino, in {\it Superconductivity}, edited by R.D. Parks (Marcel Dekker, Inc., New York, 1969).

\bibitem{McMillan_book_1969} W.L. McMillan and J.M. Rowell, in {\it Superconductivity}, edited by R.D. Parks (Marcel Dekker, Inc., New York, 1969).

\bibitem{Allen_book_1982} P.B. Allen and B. Mitrovi\'{c}, in {\it Solid State Physics}, edited by H. Ehrenreich, F. Seitz, and D. Turnbull (Academic, New York, 1982) Vol. 37.

\bibitem{Rohlf_Book_1994} 	J.W. Rohlf, {\it Modern Physics from $\alpha$ to $Z^0$} (Wiley-VCH, New-York, 1994).

\bibitem{Khasanov_Ga_PRB_2020} Rustem Khasanov, Hubertus Luetkens, Alex Amato, and Elvezio Morenzoni, {\it Structural phases of elemental Ga: Universal relations in conventional superconductors}, Phys. Rev. B {\bf 101}, 054504 (2020).\\
    \url{https://doi.org/10.1103/PhysRevB.101.054504}

\bibitem{Kittel_Book_1996} C. Kittel, {\it Introduction to Solid State Physics, 7th Ed.}, (Wiley, India, Pvt. Limited, 2007).

\bibitem{Poole_Book_2014} C. Poole, H. Farach, R. Creswick, and  R. Prozorov, {\it Superconductivity 3rd Edition} (Elseiver: Amsterdam, 2014).

\bibitem{Prakash_Science_2017} O. Prakash, A. Kumar, A. Thamizhavel, and S. Ramakrishnan, {\it Evidence for bulk superconductivity in pure bismuth single crystals at ambient pressure}, Science {\bf 355}, 52 (2017).\\
    \url{https://doi.org/10.1126/science.aaf8227}

\bibitem{Campanini_PRB_2018} D. Campanini, Z. Diao, and A. Rydh, {\it Raising the superconducting $T_{\rm c}$ of gallium: In situ characterization of the transformation of $\alpha-$Ga into $\beta-$Ga}, Phys. Rev. B {\bf 97}, 184517 (2018).\\
    \url{https://doi.org/10.1103/PhysRevB.97.184517}

\bibitem{Khasanov_Al_PRB_2021} Rustem Khasanov and Igor I. Mazin, {\it  Anomalous gap ratio in anisotropic superconductors: Aluminum under pressure}, Phys. Rev. B {\bf 103}, L060502 (2021).\\
    \url{https://doi.org/10.1103/PhysRevB.103.L060502}

\bibitem{Khasanov_Pb_PRB_2021} Rustem Khasanov, Debarchan Das, Dariusz Jakub Gawryluk, Ritu Gupta, and Charles Mielke, III, {\it Isotropic single-gap superconductivity of elemental Pb}, Phys. Rev. B {\bf 104}, L100508 (2021).\\
    \url{https://doi.org/10.1103/PhysRevB.104.L100508}

\bibitem{Leng_PRB_2019} H. Leng, J.-C. Orain, A. Amato, Y. K. Huang, and A. de Visser, {\it Type-I superconductivity in the Dirac semimetal PdTe$_2$ probed by $\mu$SR}, Phys. Rev. B {\bf 100}, 224501 (2019).\\
    \url{https://doi.org/10.1103/PhysRevB.100.224501}

\bibitem{Bussmann-Holder_EPB_2004} A. Bussmann-Holder, R. Micnas, and A. R. Bishop, {\it  Enhancements of the superconducting transition temperature within the two-band model}, Eur. Phys. J. B. {\bf 37}, 345 (2004).\\
    \url{https://doi.org/10.1140/epjb/e2004-00065-5}

\bibitem{Bussmann-Holder_Arxiv_2009} A. Bussmann-Holder, {\it Comment on: Superfluid density and specific heat within a self-consistent scheme for a two-band superconductor}, arXiv:0909.3603, unpublished.\\
    \url{https://doi.org/10.48550/arXiv.0909.3603}

\bibitem{Bussmann-Holder_CondMat_2019} A. Bussmann-Holder, H. Keller, A. Simon, and A. Bianconi, {\it Multi-Band Superconductivity and the Steep Band -- Flat Band Scenario}, Condens. Matter {\bf 4}, 91 (2019).\\
    \url{https://doi.org/10.3390/condmat4040091}

\end{thebibliography}
\end{document}